\documentstyle[psfig]{article}
\def\la{\mbox{\raisebox{-0.1ex}{$\scriptscriptstyle \stackrel{<}{\sim}$\,}}}
\def\ga{\mbox{\raisebox{-0.1ex}{$\scriptscriptstyle \stackrel{>}{\sim}$\,}}}
\newcommand{\cn}{$\overline{C_n^2} ~ $}
\newcommand{\cne}{\mbox{$\overline{C_n^2}$}}

\begin{document}

\normalsize

\title{\Large\bf Pulsar Scintillation and the Local Bubble}
\title{ }

\author{N. D. R. Bhat, Y. Gupta, and A. P. Rao}
\author{ }

\date{September 5, 1997}
\date{ }




\begin{center}
{ $ $ }
\\
\vspace{2.0cm}
{\Large\bf Pulsar Scintillation and the Local Bubble}
\\
\vspace{5.0cm}
N. D. Ramesh Bhat\footnote{email: bhatnd@ncra.tifr.res.in},
Yashwant Gupta\footnote{email: ygupta@ncra.tifr.res.in},
A. Pramesh Rao\footnote{email: pramesh@ncra.tifr.res.in}
\\
National Centre for Radio Astrophysics, Tata Institute of Fundamental Research,
\\
Post Bag 3, Ganeshkhind, Pune 411 007, India.
\\
\vspace{5.0cm}
Accepted for publication in The Astrophysical Journal
\\
\end{center}

\newpage



\begin{center}
{\bf ABSTRACT}
\end{center}

{\sf
We present here the results from an extensive scintillation study of twenty pulsars
in the dispersion measure (DM) range $ 3-35 \ {\rm pc \ cm^{-3}}$ carried out using
the Ooty Radio Telescope at 327 MHz, to investigate the distribution of ionized material in the
local interstellar medium.
Observations were made during the period January 1993 to August 1995, in which the dynamic scintillation 
spectra of these pulsars were regularly monitored over 10$-$90 epochs spanning $ \sim $ 100 days.
Reliable and accurate estimates of strengths of scattering have been deduced from the scintillation 
parameters averaged out for their long-term fluctuations arising from refractive scintillation effects.
Our analysis reveals several anomalies in the scattering strength, which suggest that
the distribution of scattering material in the solar neighborhood is not uniform.

We have modelled these anomalous scattering effects in terms of inhomogeneities in the distribution of 
electron density fluctuations in the local interstellar medium (LISM).
Our model suggests the presence of a low density bubble surrounded by a shell of much
higher density fluctuations.
We are able to put some constraints on geometrical and scattering properties of such a
structure, and find it to be morphologically similar to the Local Bubble known from
other studies.
}



{{\it Subject headings: } ISM:General -- Structure -- Bubbles -- Pulsars:General}


\vspace{1.0cm}

\section{Introduction}

The solar system is believed to reside in a very hot (temperature T $ \sim ~ 10^6 $ K)
and tenuous (electron density $ n_e ~ \approx ~ 0.005 ~ cm^{-3} $) X-ray emitting cavity, 
which is typical of the coronal phase of the ISM, possibly produced by supernovae or winds 
from hot, massive stars (Cox \& Reynolds 1987; McCammon \& Sanders 1990).
This region, often termed as the Local Bubble, has been of considerable interest to both 
observers and theorists alike owing to its proximity and atypical nature.

The X-ray data reveal that the Local Bubble has an elongated geometry, extending up to 200$-$300 pc
perpendicular to the galactic plane and up to 50$-$100 pc in the plane (Snowden et al. 1990). 
This region was also noted for its deficiency of neutral hydrogen gas (Paresce 1984; Frisch \& York 1983), and 
recent studies in the extreme-ultraviolet (EUV) waveband of a large sample of bright sources 
(Warwick et al. 1993) estimate the gas density to be $ \sim $ 0.05 $ {\rm cm ^{-3}} $.  
The local interstellar medium (LISM), which consists of the bubble and its surroundings, has 
rather loose definitions (Cox \& Reynolds 1987; Bochkarev 1987) and in this paper we use this 
term to mean a region of a few 100 pc surrounding us. Though adequate understanding exists 
to say that the LISM deviates significantly in its properties from the galactic average, 
there are several aspects that lack satisfactory understanding, specially its detailed 
morphological characteristics, accurate size, nature of the boundary and properties of
the material in it.

Propagation effects on radio waves emitted by pulsars, such as dispersion and scattering,
probe the distribution of thermal plasma in the interstellar medium (ISM).
In particular, observable effects of scattering of radio waves from pulsars, enable us to
probe the electron density fluctuations in the ISM (see Rickett (1990) for a review).
These density fluctuations are thought to arise from turbulence and hence scintillation
studies of pulsars also provide information on the nature and distribution of the interstellar 
turbulence (Cordes, Weisberg \& Boriakoff 1985). 
The distribution of electron density fluctuations in the LISM can differ significantly 
from the typical ISM due to the Local Bubble. 
It is reasonable to expect the Local Bubble and its environment to play a substantial
role in determining the scintillation properties of nearby (distance \la 1 kpc) 
pulsars.
Such pulsars therefore form potential tools to study the LISM.

Pulsar scintillation properties are best studied using their dynamic spectra $-$ records of intensity
variations in the time-frequency plane $-$ which shows random intensity modulations that fade over narrow 
frequency ranges and short time intervals. 
The phenomenon giving rise to this effect, known as diffractive scintillation, arises from scattering 
of pulsar signals by the small-scale ($ 10^7 ~ \la $ s $ \la ~ 10^9 $ m) density fluctuations present 
in the ISM (Rickett 1990; Cordes, Pidwerbetsky \& Lovelace 1986; Spangler 1988). 
The decorrelation bandwidth, {\it i.e.}, the frequency range over which the intensity decorrelates, provides 
information on the strength of scattering along the line-of-sight. 
In addition, the average scintillation properties of pulsar dynamic spectra are also expected to show 
variations over time 
scales of days to weeks at metre wavelengths as a result of refractive scintillation effects 
(Cordes et al. 1986; Romani, Narayan \& Blandford 1986; Rickett 1990).
These arise from propagation of pulsar signals 
through large-scale (s $ \ga ~ 10 ^{11} $ m) density irregularities in the ISM. 
Therefore, estimates of average scintillation properties from a few epochs of observations of dynamic 
spectra are prone to errors due to refractive scintillation effects. 
A large number of long stretches of data taken over time spans of months to years are essential 
to get reliable and accurate estimates of scintillation parameters. 
Systematic studies of this kind have not been carried out for many pulsars and most earlier
measurements were based on a few epochs of observations. 

Not much is known about the electron density distribution of the interstellar gas 
in the LISM.
Current models predict uniform electron density and uniform electron density fluctuations,
except for a smooth dependence with height above the galactic plane (Taylor \& Cordes 1993,
Cordes et al. 1985).
Recent scintillation observations of PSR B0950+08 suggest that the interior of the bubble is 
dominated by relatively lower magnitudes of density fluctuations (Phillips \& Clegg 1992).
However, there has been no accurate and systematic study of the connection between scintillation
properties of local pulsars and the structure of the LISM.

During 1993$-$95 we carried out a long-term, systematic study of the scintillation properties 
of twenty nearby pulsars, with the two-fold aim of (i) studying refractive scintillation effects 
and (ii) understanding the LISM. In this paper, we describe the results of the second study. 
The results of the first part are presented in another paper (Bhat, Rao \& Gupta 1997c, under preparation).
Our observations are described in Section 2. In Section 3, we present the results of the data 
analysis and discuss its implications on the distribution of electron density fluctuations 
in the LISM. In Section 4, we discuss the viability of simple models of specific
density structures in explaining the present observations. The success and limitations of our 
best model are briefed in Section 5, where we also compare it with pertinent results from other
published studies on the LISM. At the end, we suggest some possible tests for the present model
and some useful observations which will improve upon the present understanding of the LISM.

\section{Observations }

Pulsar observations were made using the Ooty Radio Telescope (ORT) which is an equatorially
mounted 530 m x 30 m parabolic cylinder operating at 327 MHz (Swarup et al. 1971). 
The ORT has an effective collecting area of 8000 $ m^2 $, system temperature of $ 150 ^o $ K 
and is sensitive to linearly polarized radiation with electric field in the North-South plane. 
The telescope has $ 9{1 \over 2} $ hours of hour angle coverage and a declination coverage from 
$ -55^o $ to $ +55^o $. 
The ORT is a phased array, with 1056 dipoles at its feed, signals from which are combined 
to form the signals from north and south halves. 
These signals are input to a correlation spectrometer (Subramanian 1989) 
to yield the cross power spectrum of the 
signals from the north and south halves. 
Pulsar data was taken over a bandwidth of 9 MHz. 
For pulsar scintillation observations, such spectra were obtained with 64 channels spanning
the bandwidth, yielding a frequency resolution $ \approx $ 140 kHz. 
The data were acquired both on the pulse and also on part of the off pulse regions of the pulsars. 
The data were calibrated for both the telescope gain and receiver bandpass using a continuum source 
at a declination close to that of the pulsar being studied.

Within the sky coverage and the sensitivity limits of the ORT, there are 20 nearby pulsars which 
were found suitable for studying the LISM. 
These are listed in Table 1 along with relevant observational details.
Here $ {\rm N_{ep}} $ is the number of epochs of observations and 
$ \Delta f _{ch} $ and $ \Delta t $ are the frequency and time resolutions of the dynamic spectra. 
Leaving out some faint pulsars ($ {\rm S_{400}} \la $ 25 mJy) and a few short-period pulsars 
(period $ \la $ 100 msecs), which could not be observed due to our instrumental constraints,
this list formed the complete sample of pulsars within a distance $ \la $ 1 kpc known at the time of 
our observations.
Subsequently more pulsars have been discovered and there has been a considerable increase in the 
number of local pulsars.
Our sample, despite its limited nature, has a sky coverage that is reasonably uniform (Fig. 4).
The distance estimates of pulsars are given in the column 4 of Table 1.
We adopt distance estimates based on the model for electron density distribution given by 
Taylor \& Cordes (1993).
For two pulsars, PSR B0950+08 and PSR B0823+26, there are independent distance estimates from parallax 
measurements and we adopt them in place of DM based distances.

\begin{figure}[thbp]
\centerline{ \hskip 0 cm
{\psfig{file=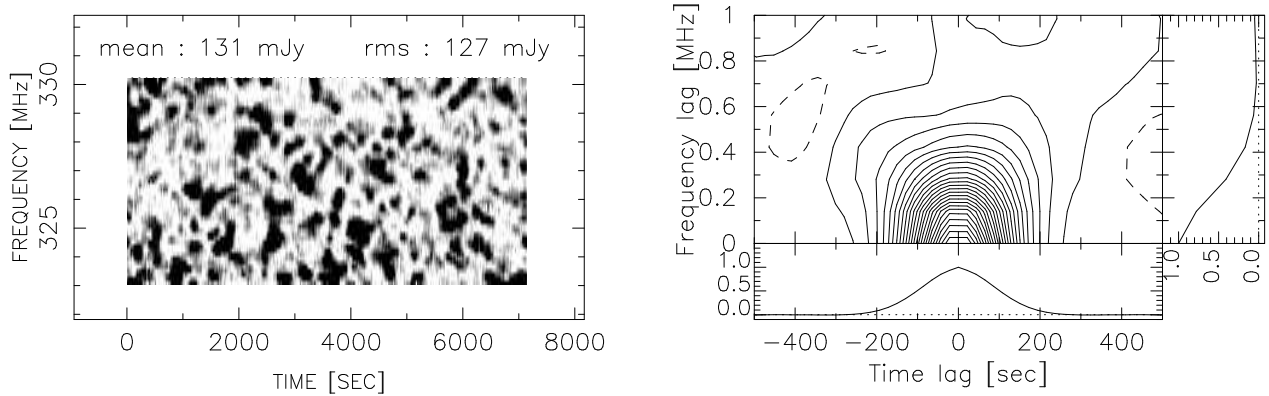,rheight=5.5cm,height=13.5cm,angle=270}}
}
\caption[]{
The dynamic scintillation spectrum of PSR B0919+06 as observed on
24 May 1994, darker areas representing higher intensity values
(left panel). The ACF is shown along with one-dimensional cuts across zero
lags of frequency and time (right panel).
}
\label{Fig01}
\end{figure}

The pulsars were observed for their dynamic scintillation spectra over a large number of epochs
(${\rm N_{ep} }\sim10-90$), spanning $ \sim $ 100 days during 1993$-$95.
The number of epochs of observations and their separations were determined by their expected 
refractive scintillation properties.
For the LISM studies, our basic aim was to obtain reliable estimates of the scintillation parameters by 
averaging out their fluctuations arising from refractive scintillation effects, which has been achieved 
except for two pulsars $-$ PSR B0950+08 and PSR B0031$-$07.
The pulsar PSR B0950+08 shows scintillation characteristics that were not measurable with our 
experimental setup (Fig. 2.a) and in the case of PSR B$0031-07$, dynamic spectra observations 
turned out to be difficult due to its nulling properties.
At every epoch of observations, the dynamic spectrum of each pulsar was recorded over a typical 
duration of 2$-$3 hours in order to get accurate estimates of the scintillation characteristics.
Further details regarding the observations and a description of our data analysis can be found 
in another paper (Bhat, Rao \& Gupta 1997a, under preparation).

Fig. 1.a shows a typical dynamic scintillation spectrum of pulsar PSR B0919+06 obtained from
observations made on 24 May 1994. 
The intensity scintillation patterns of this pulsar decorrelate over $ \sim $ 100 secs in time
and $ \sim $ 250 kHz in frequency.
Bright intensity regions, known as {\it scintles}, are resolvable when the instrumental resolutions 
in time and frequency happen to be smaller than respective decorrelation widths. 
With the frequency resolution available with our experimental setup ($ \Delta f _{ch} ~ \approx $ 140 kHz), 
it is expected to resolve the spectral features of pulsars with 
DM $ < $ 40 $ {\rm pc ~ cm ^{-3}}$ at our observing frequency. 
The required temporal resolution ($ \Delta t $), typically $ \sim $ 10 secs, is not a 
limiting factor owing to the high sensitivity of the upgraded ORT (Selvanayagam et al. 1993). 
The pulsar data was integrated for $ \sim $ 10 secs to average out the intrinsic pulse-to-pulse 
fluctuations of pulsar signals to an acceptable level. 
Our observations show that pulsar dynamic spectra vary significantly in their properties
over time scale as short as $2-3$ days, where the scintillation patterns change in terms 
of their sizes and shapes in the frequency-time plane.

Fig. 2 displays sample dynamic spectra of some selected pulsars over a DM range 3$-$10 $ 
{\rm pc ~ cm ^{-3}} $, illustrating the diversity and wide range in the scintillation 
characteristics of nearby pulsars. 
The spectra were obtained with $ \Delta f _{ch} $ $ \approx $ 140 kHz for pulsars 
PSR B1237+25, PSR B1929+10 and PSR B2327$-$20 and with $ \Delta f _{ch} $ $ \approx $ 280 kHz 
in the case of PSR B0950+08.
The temporal resolutions are in the range $ \sim $ 10$-$15 secs.
Fig. 2.a displays the spectrum of PSR B0950+08 obtained from observations made on 12 July 1995.
The scintillation patterns of this pulsar are not resolvable within our observing bandwidth and 
within typical observing durations of 2 hours, whereas the spectra of other pulsars show 
remarkable difference.
In the case of PSR B2327$-$20 (Fig. 2.d), the patterns fade over a very narrow frequency 
range ($\sim $ 100 kHz). 
Pulsars PSR B1237+25 and PSR B1929+10 show this property over a much wider frequency range 
($ \sim $ 1 MHz) (Figs. 2.b and 2.c).
Thus the scintillation properties of local pulsars span over a wide range of two orders of magnitude
at our observing frequency.

\begin{figure}[thbp]
\vskip 1cm
\centerline{ \hskip 0 cm
{\psfig{file=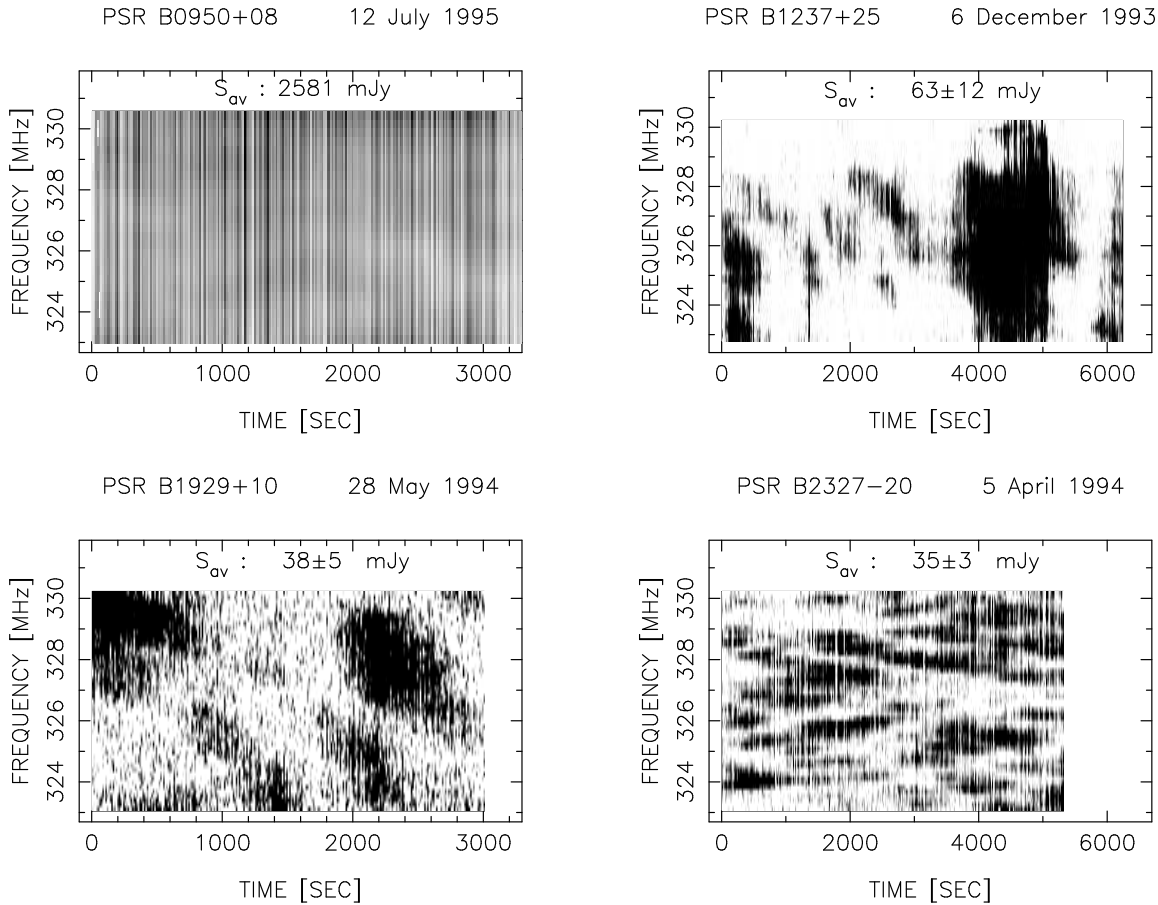,rheight=12.5cm,height=12.5cm,angle=270}}
}
\caption[]{
Sample dynamic spectra of four pulsars are shown to illustrate the wide range
of scintillation properties of nearby pulsars. The darker areas represent higher
intensity values. The white regions correspond to 20\% of the mean
intensity and the black regions to twice the mean intensity; in between, the
intensity values are linearly represented by the gray-scale. The apparent,
temporal intensity fluctuations seen for the pulsar PSR B0950+08 are at
about 25\% level from the mean, which are likely to be intrinsic variations.
$ {\rm S_{av}} $ is the average flux as measured on the observing day given
at the top of each panel.
}
\label{Fig02}
\end{figure}

\section{Data Analysis and Results }

\subsection{The 2D ACF Computation }

To quantify the average characteristics of the scintillation patterns at any epoch, we make use of 
the two-dimensional auto co-variance function (2D ACF) technique. 
The ACF was computed for a maximum frequency lag of half the observing bandwidth and for a maximum 
time lag of half the observing time. 
The function was corrected for the effect of system noise fluctuations and the residual, intrinsic 
pulse-to-pulse fluctuations, which being uncorrelated appear as a ridge-like feature at zero time 
lag in the 2D ACF. 
Fig. 1.b shows such an ACF for PSR B0919+06, for the epoch 24 May 1994, where the display has been 
restricted to a maximum time lag of 500 secs and a maximum frequency lag of 1 MHz. 
The 2D ACF can be characterized by its widths along zero frequency lag and zero time lag axes, which 
are the decorrelation bandwidth $ \nu _d$, defined as the half-maximum width along zero time lag axis, 
and the scintillation time $ \tau _d$ , defined as the $ { e ^ {-1} }$ width along zero frequency lag axis.
The 2D ACF was fitted with a two-dimensional gaussian function of the following form.

\begin{equation}
\rho _k (\nu , \tau ) ~ = ~ C_o ~ exp \left[ - ~ \left( C_1 ~ \nu ^2 ~ + ~ C_2 ~ \nu ~ \tau ~ 
+ ~ C_3 ~ \tau ^2 \right) \right]
\end{equation}

where  $ \nu $ is the frequency lag and $ \tau $ is the time lag.
The suitability of this type of function in representing the ACF of dynamic 
spectrum has been discussed by Gupta, Rickett \& Lyne (1994). 
The amplitude of the gaussian function ($C_o$) is treated as unity in our 
case since the computed ACF is normalized to unity amplitude. 
While carrying out the fitting, the deviations between the ACF and the model gaussian 
are weighted by their 
uncertainties, which are essentially the estimates of rms noise and are 
determined from the noise in the dynamic spectrum and the number of data
pairs averaged to get $ \rho _k $.

The model parameters $ C_1 $ , $ C_2 $ and $ C_3 $ are estimated by a $ \chi ^2 $ minimization procedure.
The scintillation parameters $ \nu _d $ and $ \tau _d $ can be expressed in terms of these model 
parameters as 

\begin{equation}
\nu _d ~ = ~ \left( { ln ~ 2 \over C_1 } \right) ^{0.5}
\hspace{1.0in}
\tau _d ~ = ~ \left( { 1 \over C_3 } \right) ^{0.5}
\end{equation}

The uncertainties in $ C_1 $ and $ C_3 $ are obtained from a $ \chi ^2 $ variation of unit magnitude 
from its minimum and translated into corresponding uncertainties in $ \nu _d $ and $ \tau _d $.
We also take into account the estimation errors in $ \nu _d $ and $ \tau _d $ arising due to finite number of 
scintles, given by

\begin{equation}
\sigma _{est} ~ = ~ \left( { B_{obs} ~ t_{obs} \over \nu _d ~ \tau _d } \right) ^ {-0.5}
\end{equation}

where $ \sigma _{est} $ is the fractional error, $ B_{obs} $ is the observing bandwidth and 
$ t_{obs} $ is the duration over which the dynamic spectrum was recorded. 
Errors from the model fitting are added in quadrature with the estimation errors
to get the final uncertainties in the parameters $ \nu _d $ and $ \tau _d $.
The decorrelation widths obtained in this manner are corrected for 
smearing due to finite instrumental resolutions in frequency ($ \Delta f _{ch} $) and in time
($ \Delta t $) respectively. 
Barring a few exceptions, such as $ \nu _d $ measurements of pulsars PSR B$1540-06$ and PSR B2310+42
(Table 2), the instrumental smearing is not significant for our data.
Our observations show large fluctuations of $ \nu _d $ and $ \tau _d $ and variations as large as
factor of $3-5$ are seen for most pulsars. 
The typical rms fluctuations are about $40-50$\%.
Nevertheless, our large number of epochs of observations taken over time spans of $ \sim $ 100 days
allow us to reduce the statistical uncertainties due to these fluctuations to about 5$-$10\% in the 
estimates of average scintillation parameters.

In order to get best estimates of average scintillation parameters from a given number of epochs of 
observations, we make use of a weighted average 2D ACF, which will be referred to as the Global
2D ACF (GACF).
It is obtained from our observations of dynamic spectra of the pulsar at all epochs, using the 
following definition.

\begin{equation}
\rho _g \left( \nu , \tau \right) ~ = ~ { \sum _{k=1} ^{k=N^{'}_{ep}} ~ w_k \left( \nu , \tau \right) ~ 
\rho _k \left( \nu , \tau \right) \over
{ \sum _{k=1} ^{k=N^{'}_{ep}} ~ w_k \left( \nu , \tau \right) } }
\end{equation}

where $ \rho _k $ is the ACF of dynamic spectrum at $ k^{th} $ epoch. 
$ N^{'}_{ep} $ is the number of epochs of observations made with identical resolutions in time 
and frequency. 
$ \omega _k $ is the weight function, which is simply the number of data pairs averaged to get $ \rho _k$ . 
The GACFs computed in this manner are shown in Fig. 3 for some pulsars.
The signal-to-noise ratio in these ACFs is much higher than the individual ACFs and this results in 
more reliable estimates of scintillation parameters. 
The GACF is fitted with a gaussian of the form described in equation (1) to yield parameters
$ \nu _{d,g} $ and $ \tau _{d,g}$ , which are the average decorrelation bandwidth and scintillation
time scale respectively. 
The values of $ \nu _{d,g} $ and $ \tau _{d,g} $ are presented in columns 2 and 3 of Table 2.
In the case of PSR B0950+08, The GACF method does not give any meaningful results since the
pulsar has a decorrelation bandwidth much larger than our observing bandwidth ($ \nu _d ~ \gg $ 9 MHz) 
(see Fig. 2.a).
This is consistent with the expectations based on its recent scintillation measurements at 50 MHz 
(Phillips \& Clegg 1992).

Measurements of decorrelation bandwidth and scintillation time have been reported by several 
groups earlier
(Roberts \& Ables 1982; Cordes et al. 1985; Smith \& Wright 1985; Cordes 1986; Gupta et al. 1994). 
A detailed comparison of the various measurements is discussed in our paper describing in detail 
the Ooty observations
(Bhat et al. 1997a, under preparation).
There seems to be considerable unexplained discrepancies between the various measurements of 
decorrelation bandwidth ($ \nu _d $) and scintillation time ($ \tau _d $).
Most earlier measurements were from a fewer epochs of observations and hence are prone to 
errors due to refractive scintillation effects.
On comparing our measurements with others, we find the differences to be more or less unbiased,
except with those from Cordes et al. (1985).
But we also note that measurements given in Cordes (1986) are from a more extensive data set 
and later than those reported in Cordes et al. (1985).
If we restrict ourselves to these later measurements, by and large, there seems to be reasonable
consistency between the various measurements, allowing for the fact that most earlier measurements
did not take into consideration refractive scintillation effects.
Since our measurements are from long term systematic observations, 
we have been able to obtain more reliable scintillation parameters 
by averaging out for their refractive fluctuations.
We, therefore, believe that ours are the most precise measurements made so far.

\begin{figure}[thbp]
\vskip 1 cm
\centerline{ \hskip 0 cm
{\psfig{file=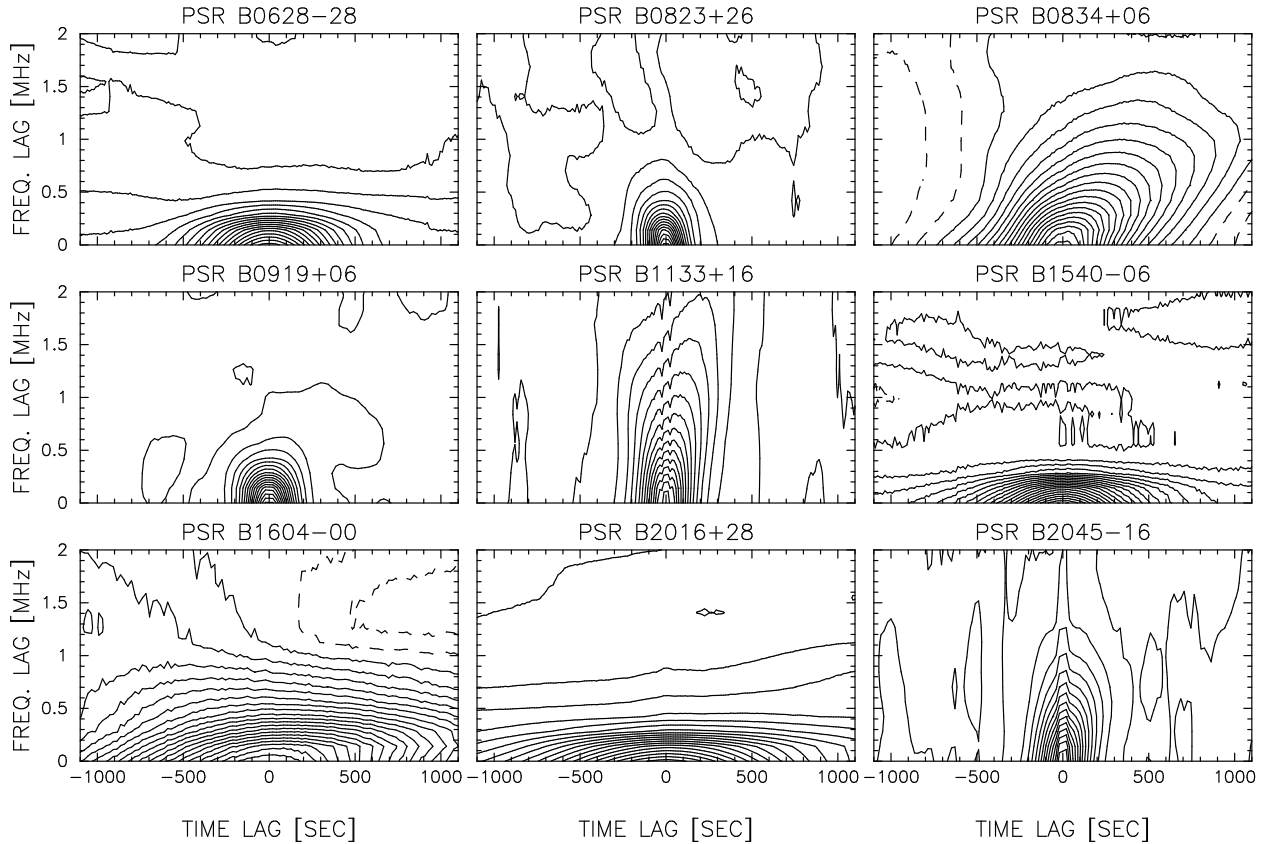,rheight=11.5cm,height=12.5cm,angle=270}}
}
\caption[]{
The Global ACFs are shown for some selected pulsars. These were computed
from $ \sim $ 10 to 30 independent ACFs. They represent the scintillation
properties averaged over the entire time spans of observations (see Table 1).
}
\label{Fig03}
\vskip 0.5 cm
\end{figure}

Among the two scintillation parameters discussed here, estimates of decorrelation bandwidths 
are better representatives of the strength of scattering.
The time scales are determined by the coherence scales as well as the transverse speeds of pulsars.
In the case of pulsars with low proper motions, the bulk motion of the medium and the earth's 
orbital motion also will have substantial roles in determining the time scales.
Furthermore, it is difficult to totally rule out intrinsic intensity variations of pulsars 
over time scales of the order of those due to ISS.
The time scales are, therefore, not as good indicators of strength of scattering as decorrelation 
bandwidths. 
We use estimates of $ \nu _{d,g} $ to derive the strengths of scattering of our pulsars.

\subsection{Estimation of Strength of Scattering}

From the estimates of decorrelation bandwidths obtained by the above-mentioned method, 
we deduce parameters characterizing the electron density fluctuations along the 
lines-of-sight of the observed pulsars. 
These density fluctuations are normally characterized by their power spectrum 
(see for example Rickett (1990)), given by

\begin{equation}
P ( \kappa ) ~ = ~ C_n^2 ~ \kappa ^{- \alpha }	
\hspace{0.5in} {2 ~ \pi \over s_o } ~ \la ~ \kappa ~ \la ~ {2 ~ \pi \over s_i }
\end{equation}

where $ \kappa $ is the spatial wavenumber, 
$ s_i $ and $ s_o $ are inner and outer cut-offs of scale size respectively. 
The amplitude of the spectrum $ C_n^2$ , which will be referred to as the strength 
of scattering in our discussion, is an indicator of the rms of electron density 
fluctuations. 
The measurable parameter decorrelation bandwidth is determined by the integral 
of $ C_n^2 $ over the line-of-sight.  
It is convenient to use a parameter  \cn , which is the line-of-sight averaged 
value of $ C_n^2 $, to characterize the scattering from different directions. 
Cordes et al. (1985) have discussed in detail the estimation of  \cn , and is 
given by 

\begin{equation}
\cne ~ = ~ A _{\alpha } ~ f _{obs} ^{\alpha } 
 ~ D ^{ - \left( { \alpha \over 2 } \right) }
 ~ \nu _d ^{ - \left( { \alpha - 2 \over 2 } \right) }
\end{equation}

where $ f _{obs} $ is the observing frequency in MHz, D is the distance estimate of the pulsar in pc and 
$ A_{\alpha } $ is a model dependent constant. 
\cn is expressed in its standard units of $ m ^{-20/3}$. 
Armstrong, Rickett \& Spangler (1995) have shown that the electron density power spectrum in the local ($ \la $ 1 kpc) 
interstellar medium can be best represented by a power-law model, quite close to the 
{\it Kolmogorov } form. 
Therefore, we assume the Kolmogorov power-law index $ \alpha = { 11 \over 3 } $ in our calculations. 
In this case, $ A _{\alpha } = 2 \times 10 ^ {-6} $ for the units used in our analysis. 
Our estimates of  \cn are given in the column 4 of Table 2.
The uncertainties shown here are based on the error estimates of $ \nu _{d,g} $ values.
For a uniformly distributed scattering medium,  \cn is expected to be a constant.

Our derived values of  \cn (Table 2) show roughly two orders of magnitude fluctuations, 
ranging from $ 10 ^ {-4.8 \pm 0.02} $ to $ 10 ^ {-3.1 \pm 0.02} $  $ {\rm m ^ { - 20 / 3 }} $.
For PSR B0950+08, we have only an upper limit on  \cn 
($ \ll 10 ^ {-3.7 } ~ {\rm m ^ {-20/3}} )$ due to our observing bandwidth limitations.
This is consistent with its recent scintillation measurements at 50 MHz (Phillips \& Clegg 1992), 
where a value  of $ \approx ~ 10^{-4.5 \pm 0.1} ~ {\rm m ^ {-20/3}} $ was reported.
Note that, for pulsars PSR B$1747-46$, PSR B2310+42 and PSR B$2327-20$,  \cn 
estimates have been obtained for the first time.

Cordes et al. (1985) measured  \cn for 15 pulsars in our sample.
Their results, however, differ considerably from those obtained from our observations.
We find that, barring a few exceptions, their  \cn values are, in general, 
larger than our values. 
Even after correcting their estimates for the new pulsar distances (Table 2) considerable
discrepancy is seen for 10 pulsars.
Reasonable agreement (within a difference of 10$-$30\%) is seen with 5 pulsars,
whereas discrepancies ranging over 2$-$10 times is the case with the remaining ones, 
for which the revised estimates continue to be larger than ours.
For pulsars PSR B0919+06 and PSR B1919+21, our estimates are about 10 times lower. 
In the case of PSR B0950+08, we have only an upper limit on  \cn, which is 
6 times lower, and a much lower value (60 times) was reported by Phillips \& Clegg (1992). 
Thus the  \cn measurements of Cordes et al. (1985) are, by and large, in 
disagreement with those from our observations. 
Since our  \cn estimates have been derived using the measurements obtained from a large number of long 
stretches of data taken over time spans $ \sim $ months to years, they are not prone to 
errors due to refractive scintillation effects. 
The $ \nu _d $ measurements used in our calculations are known with uncertainties as low as 
$ \sim $ 5$-$10\%. 

\subsection{Distribution of  \cn in the LISM}

Cordes et al. (1988) and Cordes et al. (1991) 
studied the distribution of  \cn in the galaxy by combining 
scintillation measurements of a large number of pulsars and radio sources. 
The analysis of Cordes et al. (1988) revealed large fluctuations, of several orders of magnitude, 
of \cn and the statistics vary strongly with galactic latitude.
Cordes et al. (1991) give a more refined distribution, but the details such as an 
additional inner galaxy component is not relevant as far as the LISM is concerned.
The empirical models given by them for the variation of the diffuse component of  \cn
have an exponential distribution with z-height and a gaussian in galactocentric radius
(eq. [6] of Cordes et al. (1988)).
As per their models, expected fluctuations over a region $ \la $ 1 kpc around the Sun is $ \sim $ 
3$-$6 times, whereas our observations show much larger fluctuations, about two orders of 
magnitude.
This implies that, even within the LISM, scattering material is rather non-uniformly distributed, 
and the empirical models given by Cordes et al. (1988) and Cordes et al. (1991) are too simplistic
for the LISM.

\begin{figure}[thbp]
\centerline{ \hskip 0 cm
{\psfig{file=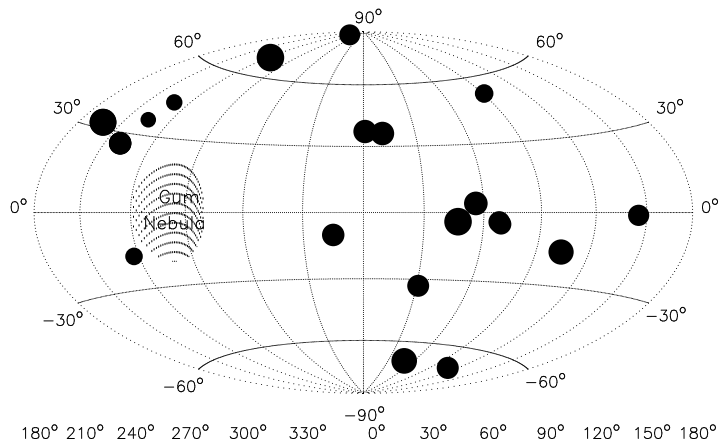,rheight=7.0 cm,height=18.5cm,angle=0}}
}
\caption[]{
The sky distribution of $ {\rm \overline { C_n^2 } } $ in the galactic
co-ordinate system, illustrating its observed fluctuations in the LISM.
The size of the black circle represents log  $ {\rm \overline { C_n^2 } } $.
The hatched region around ($ l ~ \approx ~ 260^o $ , $ b ~ \approx ~ 0^o $)
is the direction of the Gum Nebula, located at a distance of $ \sim $ 500 pc.
}
\label{Fig04}
\end{figure}

\begin{figure}[thbp]
\centerline{ \hskip 0 cm
{\psfig{file=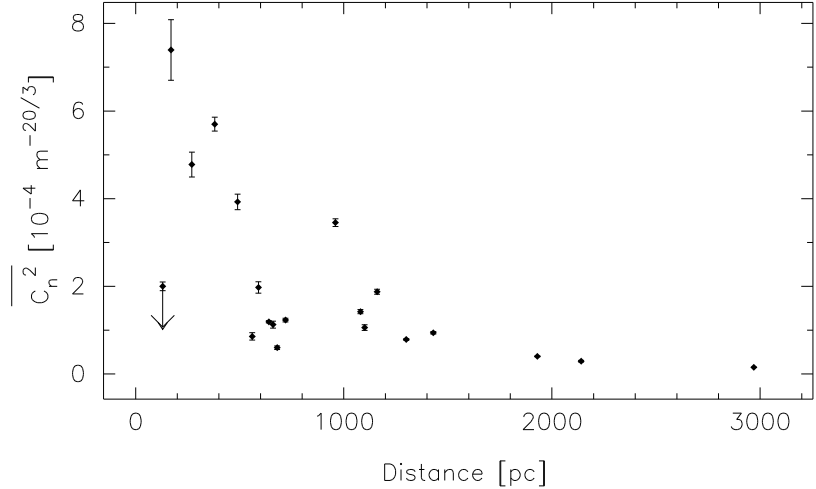,rheight=6.5 cm,height=6.5cm,angle=0}}
}
\caption[]{
The plot of $ {\rm \overline { C_n^2 } } $
against pulsar distance estimates. The upper limit shown by the downward arrow
is for the pulsar PSR B0950+08.
}
\label{Fig05}
\end{figure}

The observed fluctuations of our  \cn measurements is shown in Fig. 4. 
A complex distribution of this kind for the LISM has not been reported before.
Our pulsar sample provides a fairly uniform coverage of the sky. 
Nevertheless, no large scale trends are apparent for  \cn ($l,b$) and there are
no signatures of any clustering that is confined over a latitude or a longitude range.
Also, note that our sample does not consist of any pulsar whose line-of-sight passes 
through the Gum Nebula and hence there is no need to consider a possible enhancement 
of  \cn due to the same.

Our estimates of  \cn are plotted against pulsar distance estimates in Fig. 5. 
A systematic trend is evident for the variation of  \cn with distance.
There are significant enhancements of  \cn values within
$ \sim $ 1 kpc 
surrounding the Sun, implying a probable local origin.
Although a firm statement cannot be made, 
the  \cn values appear to converge to a steady value 
at larger distances ($ \ga $ 2 kpc). 

\begin{figure}[thbp]
\vskip -0.75 cm
\centerline{ \hskip 0 cm
{\psfig{file=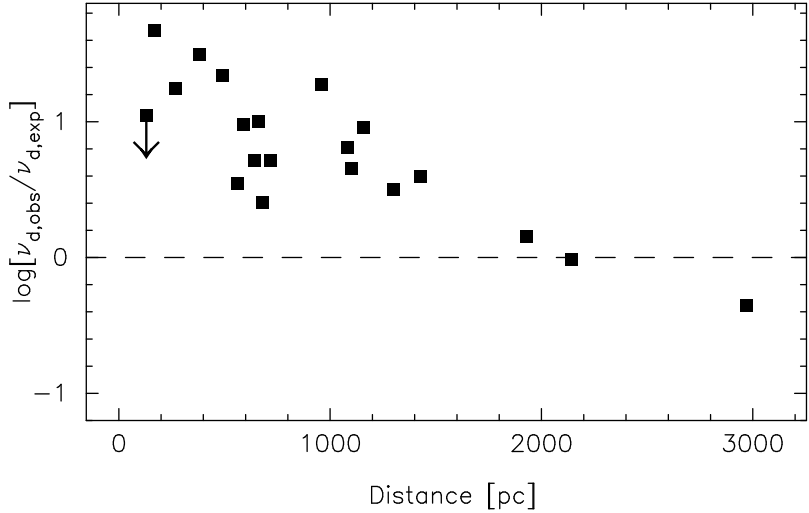,rheight=8.5 cm,height=18.5cm,angle=0}}
}
\caption[]{
The deviations of the observed decorrelation bandwidth ($ \nu _{d,obs} $) measurements
from their values expected ($ \nu _{d,exp} $) in the case of a uniform scattering medium
with $ {\rm \overline { C_n^2 } } ~ \approx ~ 0.2 \times 10 ^ {-4} ~ {\rm m ^ {-20/3}} $.
The upper limit (downward arrow) is for PSR B0950+08.
}
\label{Fig06}
\vskip 1 cm
\end{figure}

The behaviour of \cn with distance, DM and $(l,b)$, have been investigated earlier to understand 
the nature of distribution of turbulence in the galaxy.
Cordes et al. (1985) made \cn measurements of a large number of pulsars and their data showed
fluctuations as large as 5 orders of magnitude over a DM range $3-475$ $ {\rm pc ~ cm ^ {-3}} $
or distances up to $ \sim $ 10 kpc (Fig. 9 in Cordes et al. (1985)). 
If we examine only those pulsars which are closer than 1 kpc (similar to our sample), 
this data also show a similar, though weaker, trend with distance.
However, the effect was not noticed since their emphasis was on studying large-scale galactic distribution.
The effect might have got further subdued as their measurements were prone to errors due to refractive 
scintillation effects.
Subsequent to Cordes et al. (1985), pulsar distances have been revised (Taylor \& Cordes 1993).
We have examined the above data with revised \cn and the weak trend remains unchanged.
Since our estimates of \cn are obtained from more precise measurements of $ \nu _d $, 
where long term refractive fluctuations have been averaged out, we see a more obvious trend in Fig. 5.

To further assess the necessity of a non-uniform local scattering medium to explain the present 
observations, we studied the deviations of the present scintillation measurements relative 
to those expected from a uniform medium. 
We assumed a uniform medium with  \cn $ \approx ~ 0.2 \times 10 ^ {-4} ~ {\rm m ^{-20/3}}$,
which is the average of  \cn of pulsars at $ \sim $ 2$-$3 kpc where the observed trend
is closer to that of a uniform medium. 
The discrepancies between the observations and the expectations are shown in terms of ratios of 
decorrelation bandwidths (Fig. 6).
Observed measurements show considerable deviations from those due to a uniform medium 
and a systematic trend with distance, similar to the case of  \cn, is seen.
Most nearby pulsars show enhanced scattering relative to a uniform medium. 
The wide ranges and systematic trends of the observables $ \nu _d $ and the derived parameter 
 \cn suggest a non-uniform, but organized distribution 
of scattering material in the LISM.

\subsection{Anomalies in Strengths of Scattering}
\bigskip
Our observations reveal several cases of pulsars with similar DMs or at similar distances showing 
remarkably different scintillation characteristics. 
Pulsars PSR B0950+08 and PSR B1929+10 form one of the prominent examples from our data (Figs. 2.a and 2.b), 
where despite having similar DMs and comparable distances (Table 1), their decorrelation bandwidths differ 
by 1$-$2 orders of magnitude. 
Such effects will be referred to as scattering anomalies in our discussion. 
Pulsars PSR B1237+25 and PSR B$2327-20$ are another example showing this property (Figs. 2.c and 2.d)
where their decorrelation bandwidths differ by a factor of four, though they are of similar DMs. 
In order to quantify such effects, we use the following method.

If we consider two pulsars such that their DMs differ by less than 10\%, then their scattering strengths 
are also expected to be similar in the case of a uniform scattering medium. 
To quantify the deviation from such behaviour, 
we define an anomaly parameter ($ A_{dm} $) as

\begin{equation}
A_{dm} ~ = ~ { \left( \nu _{d_1} / \nu _{d_2} \right) _{observed} \over 
\left( \nu _{d_1} / \nu _{d_2} \right) _{expected} }
\end{equation}

where $ \nu _{d_1} $ and $ \nu _{d_2} $ are the decorrelation bandwidths of pulsars compared for anomaly. 
The expected values are for a given model of the scattering medium. 
In the absence of any prior information, we compute the expected values for a uniform scattering medium.
For this, we assume that the power spectrum of density fluctuations has 
a Kolmogorov form and consider standard thin-screen scattering models. 
We also assume that the dependence of scattering with DM is in a similar form to that of distance. 
Since the parameter $ A_{dm} $ quantifies the relative anomaly, 
it also has the advantages that it is less sensitive to any 
model dependent constants related to assumptions made about scattering screen geometries. 
We define this parameter to be always above unity.

\begin{equation}
ie, {\rm if} \hspace{0.5cm} A_{dm} ~ < ~ 1, \hspace{0.5cm} 
{\rm then} \hspace{0.5cm} A_{dm} ~ = ~ {1 \over A_{dm}}
\end{equation}

Though we loose the information regarding the sense of the anomaly here, it has the advantage that
when averaged over different directions of the sky, anomalies of opposite sense do not cancel each other. 
This means the parameter will have significant values only when scattering anomalies are consistently 
present in several directions. 
The parameter will have a unity magnitude in the cases of a uniform medium or spherically symmetric,
non-uniform medium in the solar neighborhood. 
The $ A _{dm} $ values are binned over a DM interval of 3 $ {\rm pc ~ cm ^ {-3}} $ to compute 
the average anomalies.
The bins are overlapped to include all possible pairs of pulsars in the analysis.
A plot of the anomaly parameter against DM is shown in Fig. 7.a, 
where the expected values are calculated for a uniform scattering medium with  \cn
$ \approx ~ 0.2 \times 10^{-4} {\rm m ^ {-20/3}} $.
It shows a systematic variation where 
the strength of anomaly decreases with DM. 
The parameter converges to a value quite close to unity at 
about DM $ \sim $ 13 $ { \rm pc ~ cm ^{-3} } $.

\begin{figure}[thbp]
\vskip -3cm
\centerline{ \hskip 0 cm
{\psfig{file=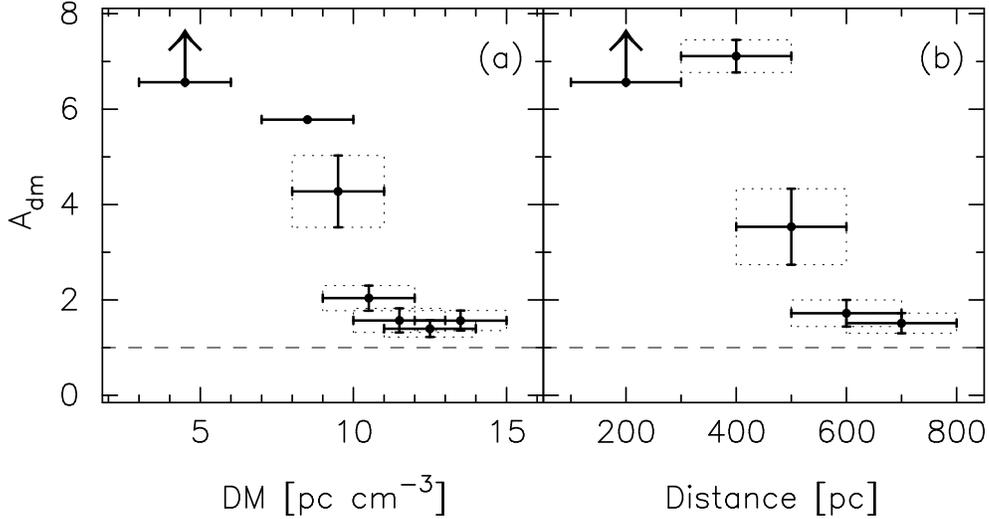,rheight=10.5cm,height=17.5cm,angle=270}}
}
\caption[]{
The anomaly curves - plots of the anomaly parameter $ A _{dm} $ computed for
similar DMs (a) and for similar distances (b) - are shown.
The dashed line is the expected curve in the cases of a uniform medium
or a spherically symmetric density structure for the solar neighborhood.
}
\label{Fig07}
\end{figure}

A similar anomaly parameter can be defined for pulsars with comparable distances. 
Because of the relatively larger uncertainties involved with distances, we consider pulsars 
differing by $ \la $ 30\% in their distance estimates for comparison for anomalous scattering. 
The anomaly values are binned over a distance range of 200 pc and overlapped so as to include 
all possible pairs of pulsars.
The anomaly parameter is plotted against distance in Fig. 7.b. 
Like in the case of DM, a systematic trend is seen with increasing distance and the magnitude of anomaly 
converges to a value close to unity at about D $ \sim $ 700 pc. 

The main implications of these anomaly curves are the following:

(i) The relative anomaly technique we used would not have given any meaningful results in the
case of spherically symmetric density structures for the solar neighborhood. The anomaly curves,
therefore, imply a structure that is highly asymmetric relative to the location of the Sun.

(ii) The strengths of anomalies are considerably larger at lower DMs ($ \la $ 10 $ {\rm pc ~ cm ^ {-3}} $)
and at smaller distances ($ \la $ 600 pc) and converge to values quite close to unity at higher 
values of DM or D. This re-affirms a local origin for the scattering anomalies.

(iii) Since the anomalous scattering effects, to a first order, are confined within a region 
$ \la $ 500 pc, they are likely to be connected to a large-scale galactic structure such as 
the Local Bubble.

\section{Modelling the Structure of the LISM}

We try to explain our observations in terms of inhomogeneities in the distribution of electron 
density fluctuations in the LISM. 
We assume that these density fluctuations are distributed in the form of a large scale, coherent structure. 
Our model has to account for 
(i) the enhanced scattering observed for nearby pulsars (Fig. 6) and 
(ii) scattering anomalies and their observed systematic trends with DM and distance (Fig. 7). 
We consider different types of density structures and examine the viability of each of them in explaining 
the present observations. 
Due to the limited number of measurements, we restrict ourselves to the simplest possible models with 
as few free parameters as possible. 

The model has to be specified by parameters characterizing its size, location and
density fluctuations. 
We model the structure as a simple ellipsoid, the size of which is parameterized in terms of three 
semi-major axes a, b and c (Fig. 11) where a and b are in the galactic plane and c is in a plane 
perpendicular to the galactic plane. 
The centre of the structure is at an offset $ r_c $ from the Sun towards the direction $ ( l_c, ~ b_c ) $.
The scattering strengths are specified by $ C_n^2 $ values. 
The number of $ C_n^2 $ parameters would vary depending upon the number of distinct scattering 
components in the model. 
Even the simplest possible model will have to be specified by a minimum of 8 free parameters
and it may not be possible to uniquely determine all of them from the present observations.

A $ \chi ^2 $-analysis to determine the best fit parameters of the model is not practicable
due to the limited number of measurements. 
Therefore we have followed an optimization procedure in which the parameters of the model 
are adjusted to get the best agreement with the observations.
We first assume some initial values for the model parameters largely based on X-ray, UV 
and HI studies of the LISM. 
Scattering strengths ($ C_n^2 $ values) are assumed depending on the nature of the model 
considered.
We then estimate predicted values of $ \nu _d $ for the observed pulsars
using the method described below.
From these, the anomalies expected for the model ($ A ^m _{dm} $) and the ratios of the 
observed decorrelation bandwidths ($ \nu _{d,obs} $) to their predicted values ($\nu _{d,exp} $) 
are computed to examine the agreement between the observations and the model predictions.
We introduce two quantities $ \epsilon _A $ and $ \epsilon _B $, that are similar to 
$ \chi ^2 $, as quantitative measures of the agreement, that are given by 

\begin{equation}
\epsilon _A ~ = ~ { 1 \over N_A } ~ \sum \left[ {\rm log} \left( { A ^m _{dm} } \right) \right] ^2
\end{equation}

\begin{equation}
\epsilon _B ~ = ~ { 1 \over N_P } ~ \sum \left[ {\rm log} \left( { \nu _{d,obs} \over \nu _{d,exp} } \right) \right] ^2
\end{equation}

where $ N_A $ is the number of anomaly ratios and $ N_P $ is the number of pulsars.
The log scale has been chosen so as to give equal weights to discrepancies that are below 
and above unity while computing $ \epsilon _A $ and $ \epsilon _B $.
In the first iteration, the free parameters are individually varied over a wide range 
(0.1 to 5 times the initial values for a, b, c, $r_c$, $ C ^2 _{n,in} $ and $ C ^2 _{n,out} $,
$0^o - 360 ^o $ for $ l_c $ and $ -90 ^o - 90 ^o $ for $ b_c $) and are set to values for which 
$ \epsilon _A $ and $ \epsilon _B $ are found to be minimum. 
The above procedure is repeated iteratively till the quantities $ \epsilon _A $ and $ \epsilon _B $ reach
minimum values that do not vary significantly. 
For each iteration, the starting values for the parameters used are the results from the previous 
iteration and the parameters are varied on a finer grid than the previous iteration.

To compute the predicted decorrelation bandwidth from our multi-component scattering model, we 
use the following technique. 
First, we note that for a homogeneous scattering medium, the decorrelation bandwidth is related to
the strength of scattering by the relation

\begin{equation}
\nu _d ~ \propto ~ { 1 \over D } ~ \left( \int _0 ^D C_n^2 (z) ~ dz \right) ^ { 2 \over 2 - \alpha }
\end{equation}

where the proportionality constant is $ ( A _{\alpha } ~ f _{obs} ^ {\alpha } ) ^ {-1} $. 
Our models can be treated as inhomogeneous media, consisting of multiple components of different
scattering strengths located at different points along the lines-of-sight.
Since the decorrelation bandwidth is essentially determined by path length differences of scattered rays, 
contributions from $ C _n ^2 (z) $ to it need to be appropriately weighted such that scattering 
regions near the source or the observer produce less path length differences than those that are midway.
Cordes et al. (1985) have discussed this aspect in detail and following their work 
(eq. [8] of Cordes et al. (1985)), for the case of a multi-component medium, equation (11) 
can be rewritten as 

\begin{equation}
\nu _d ~ \propto ~ { 1 \over D } ~
\left( \int _0 ^{D_1} { z \over D } \left( 1-{ z \over D } \right) ~ C_{n,1} ^2 (z) ~ dz
~ + ~ \int _{D_1} ^{D_2} { z \over D } \left( 1-{ z \over D } \right) ~ C_{n,2} ^2 (z) ~ dz
 ~ + ~ \cdots ~ \right) ^ { 2 \over 2 - \alpha }
\end{equation}

where the subscripts 1, 2, etc. on $ C_n^2 $ denote different components, with the medium 1 extending from 
0 to $ D_1 $, the medium 2 from $ D_1 $ to $ D_2 $ and so on.
$ D_1 $, $ D_2 $, etc. are computed along the lines-of-sight of the observed pulsars for a given geometry of
the model and $ \nu _d $ values are estimated by using the model values of $ C_{n,1} ^2 $, $ C_{n,2} ^2 $, 
etc., from which we compute the anomaly ratios as discussed earlier.
The electron density spectrum in the ISM is generally believed to be Kolmogorov.
The observations available at present cannot estimate the spectral shape in different components of the ISM.
In the absence of any other information, we assume a Kolmogorov density spectrum for all components, and 
various components of our model are characterized by their respective $ C_n^2 $ values.

\subsection{Two-component Models}

We have considered two classes of models viz., two-component and three-component models. 
Different models that are treated by us under these two classes are listed in Table 3. 
Since it is believed that the solar system resides in a low density cavity (Cox \& Reynolds 1987), called
the Local Bubble, the simplest structure we have considered is a `cavity', where the solar 
neighborhood has a deficiency of scattering material ({\it i.e.}, lower magnitudes of electron
density fluctuations) compared to the ambient ISM. 
This model is consistent with the general belief that rms electron density $ \Delta n_e ~ \propto ~ n_e $. 
This is referred to as the Model I(a).
It is basically a two-component model, scattering geometry of which is as shown in Fig. 8.a.
The density fluctuations (or equivalently the strengths of scattering) of the inner and the outer media are 
represented by $ C ^2 _{n,in} $ and $ C ^2 _{n,out} $.
The ambient medium is considered as the normal ISM, which is known to have a strength of scattering that
varies with z-height (Cordes et al. 1988) and we adopt a scale height $ z_o ~ \approx ~ 500 $ pc for 
$ C ^2 _{n,out} (z) $.
We assume that $ C_n^2 $ is uniformly distributed inside the cavity, in the absence of any prior 
information regarding the same.
The optimization procedure as described earlier has been carried out and we find this model fails to 
reproduce the observed scattering anomalies and does not give rise to enhanced scattering strengths
for nearby pulsars (Figs. 9.c, 9.d and 10.a).


Although it is generally assumed that $ \Delta n_e ~ \propto ~ n_e $, the exact relation is not obvious
and there can be exceptions to this.
Therefore the possibility of existence of a low density region with larger magnitudes of 
electron density fluctuations cannot be totally ruled out.
We have investigated such a scattering structure (referred to as Model I(b) in Table 3),
in which the solar system is embedded in a region of relatively larger magnitude of density 
fluctuations than the ambient ISM ($ C^2 _{n,in} > C ^2 _{n,out} $). 
Such a structure, however, would fail explain the low scattering strength 
observed with closest pulsar PSR B0950+08. 
Since there are not many pulsars with low scattering strengths, we examined the viability of this model 
as well. 
The model, however, did not reproduce the observed anomalies and its comparison with observations is 
shown in Figs. 9.e, 9.f and 10.b.

\begin{figure}[thbp]
\vskip 1cm
\centerline{ \hskip -9cm
{\psfig{file=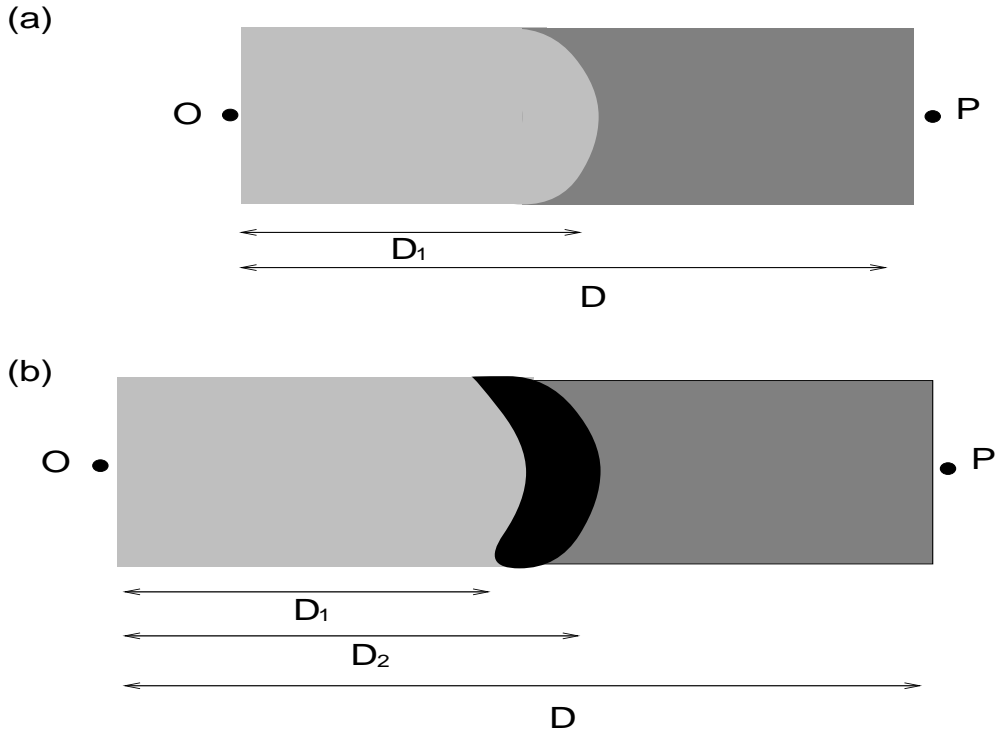,rheight=9.50cm,rwidth=5.0cm,height=9.5cm,width=13.0cm,angle=270}}
}
\caption[]{
Schematics showing the scattering geometries of different models.
O is the location of the observer and P is that of pulsar.
The distribution of $ {\rm C_n^2 } $ along the line-of-sight is shown through the
gray-scale representation, where darker regions correspond to larger
values. (a) is for Model I(a) and (b) is for Model II(a).
}
\label{Fig08}
\end{figure}

\begin{figure}[thbp]
\centerline{ \hskip -9cm
{\psfig{file=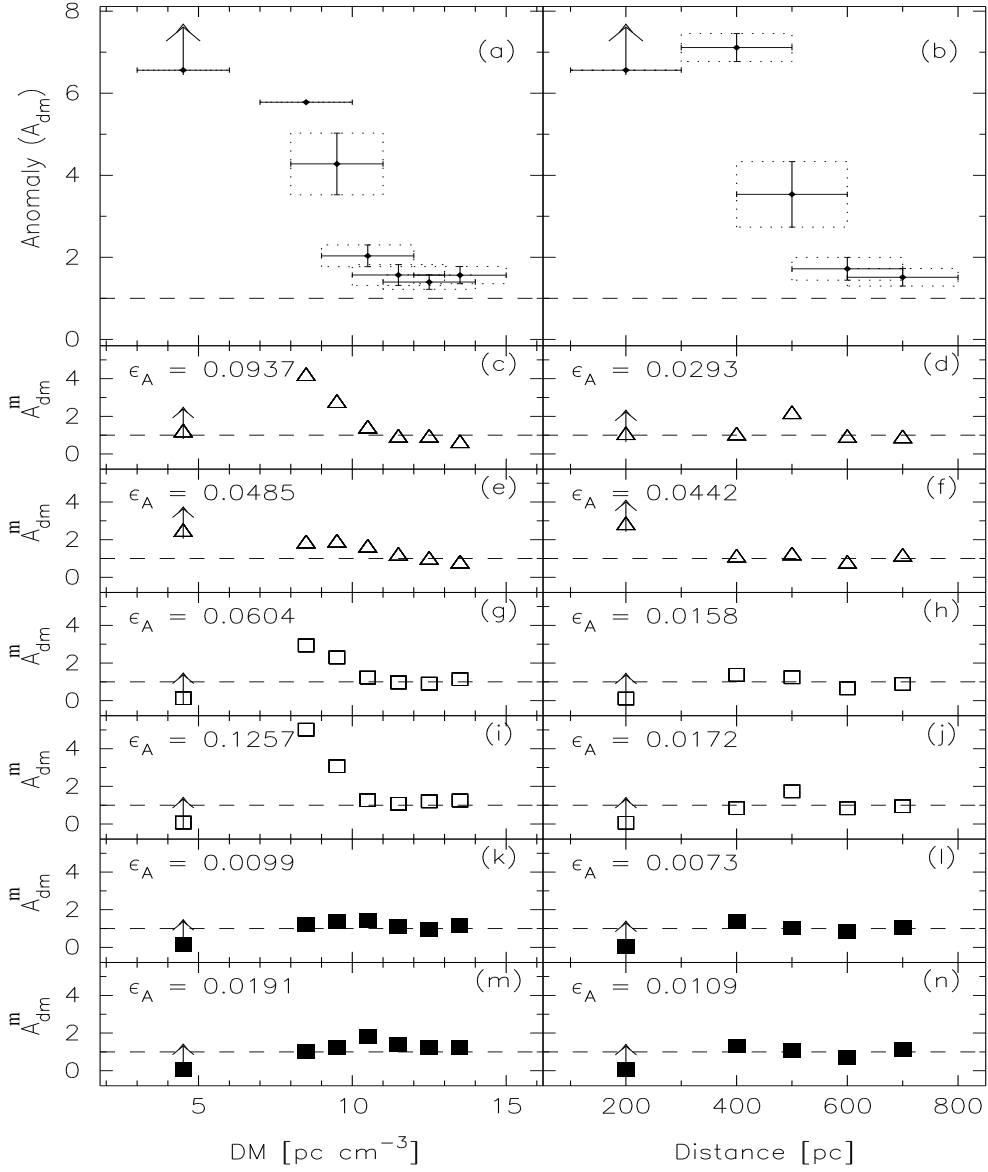,rheight=15.5cm,rwidth=5.0cm,height=15.5cm,width=13.0cm,angle=0}}
}
\caption[]{
The agreement between the observations and the predictions of various models are shown
in terms of plots of their ${\rm A_{dm}}$ values against the DMs and the distances.
The top panels (a and b) are for a uniform scattering medium.
(c) and (d) are for Model I(a) and (e) and (f) for Model I(b).
Panels (g)$-$(j) are for Model II(a) for the two possible geometries shown in Fig. 11
and (k)$-$(n) are for Model II(b), for the two geometries.
}
\label{Fig09}
\end{figure}

\begin{figure}[thbp]
\centerline{ \hskip -8cm
{\psfig{file=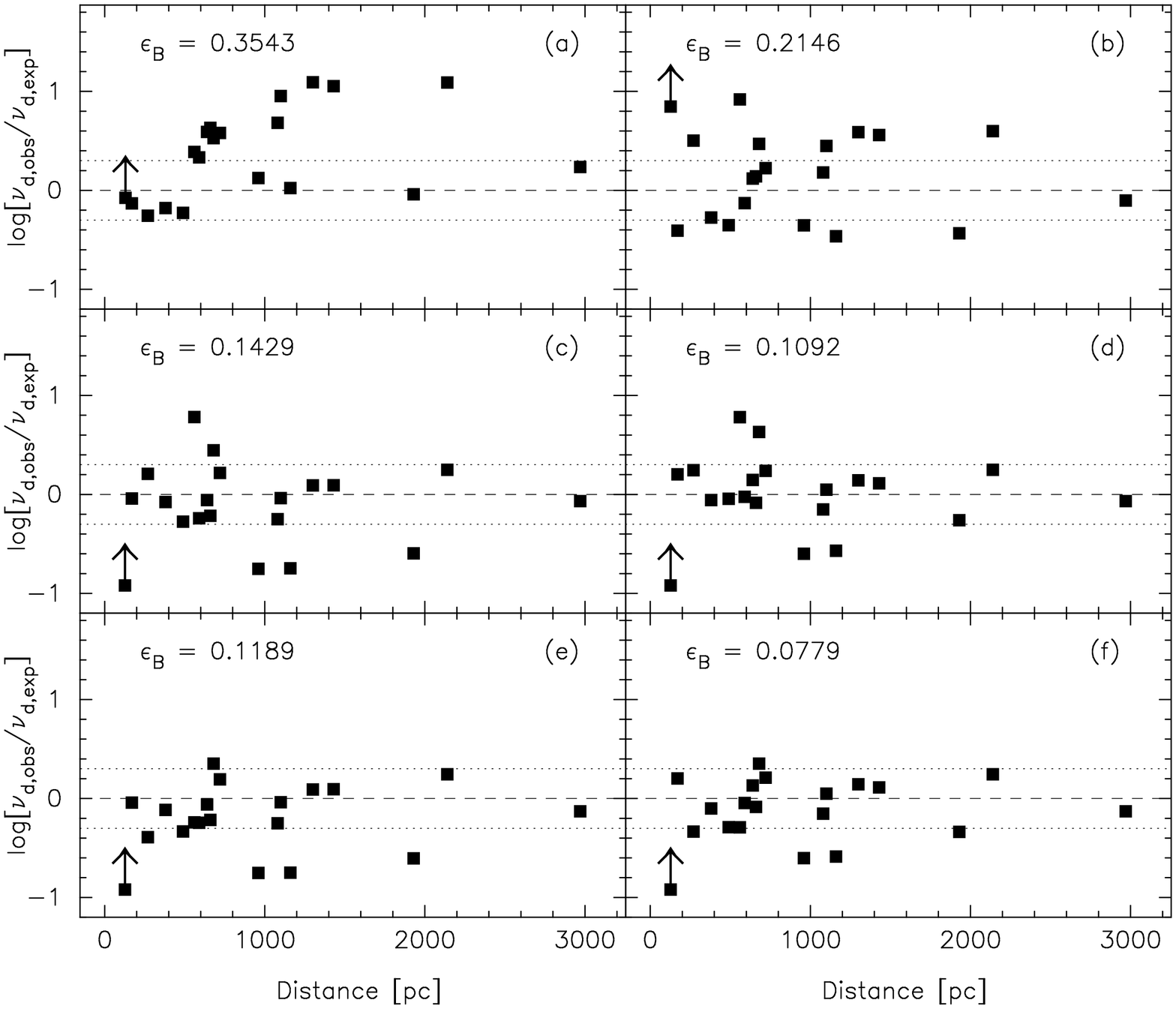,rheight=12.50cm,rwidth=5.0cm,height=12.5cm,width=13.0cm,angle=270}}
}
\caption[]{
The ratios of the observed decorrelation bandwidths ($ \nu _{d,obs} $) to their
values predicted by various scattering models ($ \nu _{d,exp} $) are shown.
(a) is for Model I(a) and (b) for Model I(b).
(c) and (d) are for Model II(a), for the solid and dashed geometries of Fig. 11
respectively.
(e) and (f) are for Model II(b), for the two geometries.
The dotted lines correspond to a discrepancy of a factor of two.
}
\label{Fig10}
\end{figure}

\subsection{Three-component Models}

The second class of models we have considered are three-component models, where we have added to the Model
I(a) a shell of enhanced scattering.
Such a structure is plausible since supernova-produced bubbles are expected to have dense shells 
surrounding them. 
Such a structure (referred to as Model II(a) in Table 3)
can give rise to both the low scattering strength for the closest pulsar
and enhanced scattering strengths for nearby pulsars. 
Enhanced scattering can be expected for pulsars outside
the shell boundary.
This will have a systematic decreasing trend with distance
since the relative contribution from the shell to the total scattering decreases with distance. 
To parameterize this model, in addition to the parameters described above, we also need to consider 
parameters characterizing scattering from the shell. 
To simplify the modelling procedure, we consider the case of a `thin' shell, with a thickness much smaller 
than pulsar distances ($ d \ll D $). 
The scattering strength due to the shell material is specified by its density fluctuation parameter 
$ C ^2 _{n,sh} $ and we assume that density fluctuations have a uniform distribution within the 
shell region. 
To simplify the model, we characterize the scattering from the shell by the integral of $ C_{n,sh} ^2 $ 
over the thickness, thereby requiring only one free parameter for the shell. 
We also assume that this dense shell has much higher levels of density fluctuations than both the interior 
of the bubble and the ambient ISM. 
The scattering geometry in this case is as shown in Fig. 8.b.
We find the shell structure succeeds to some extent in 
reproducing some of the observed scattering anomalies and we get a reasonable agreement with the observed
anomaly curves (Figs. 9.g, 9.h and 10.d). 
Though there is significant improvement compared to earlier cases of simpler structures, as can be seen from 
their $ \epsilon _A $ and $ \epsilon _B $ values, there exists some disagreement which can be seen as 
significant deviations of some of the anomaly values from unity (Figs. 9.g to 9.j) and discrepancies in
$ \nu _d $ values of some pulsars (Figs. 10.c and 10.d).

On carefully examining the predictions of Model II(a), we find the agreement with the observations is poor in 
the case of pulsars at larger z-heights. 
To account for this, we assume that the strength of scattering from the shell decreases with $|z|$, 
like in the case of ambient ISM. 
This is a variant of the previous model and is called Model II(b).
If we presume that both the cavity and its shell boundary have formed out of rather smoothly distributed 
galactic disk component, it is reasonable to assume that the density fluctuations in the shell have
characteristics 
similar to that of density variations of the disk component. 
We assume a scale height $ z_d ~ \sim ~ 135 $ pc for $ C ^2 _{n,sh} (z) $, 
which is the scale height of the neutral gas density variations in the galactic disk component (Bloemen 1987).
This would lead to substantial changes in the magnitude of density fluctuations with $ | z | $ and 
can give rise to relatively weaker scattering strengths for pulsars at larger z-heights than
that is caused by the earlier model.
We carried out the optimization procedure to determine the best parameters of this model and find that 
it reproduces the observed anomalies quite well.
The excess scattering ($ \nu _d $ ratios) is also better accounted for.
The agreement between the observed and predicted anomalies is shown in Figs. 9.k, 9.l and 10.f 
and is the best among different models considered by us. 
Note that the anomaly values $ A ^m _{dm} $ and $ \nu _{d,obs} / \nu _{d,exp} $
are quite close to unity, suggesting the 
success of the model in explaining the present observations. 
In addition, residual anomalies are randomly placed with respect to zero, thereby ruling out any 
systematic trends.
The geometrical parameters and the scattering properties of the local 
scattering structure inferred from the model are listed in Table 4.
The suggested geometry is schematically shown in Fig. 11 (the solid curve), as cuts through the 
galactic plane (Fig. 11.a) and along a plane perpendicular to $ b ~ = ~ 0^o $ and passing 
through the galactic poles (Fig. 11.b). 

\begin{figure}[thbp]
\centerline{ \hskip -1 cm
{\psfig{file=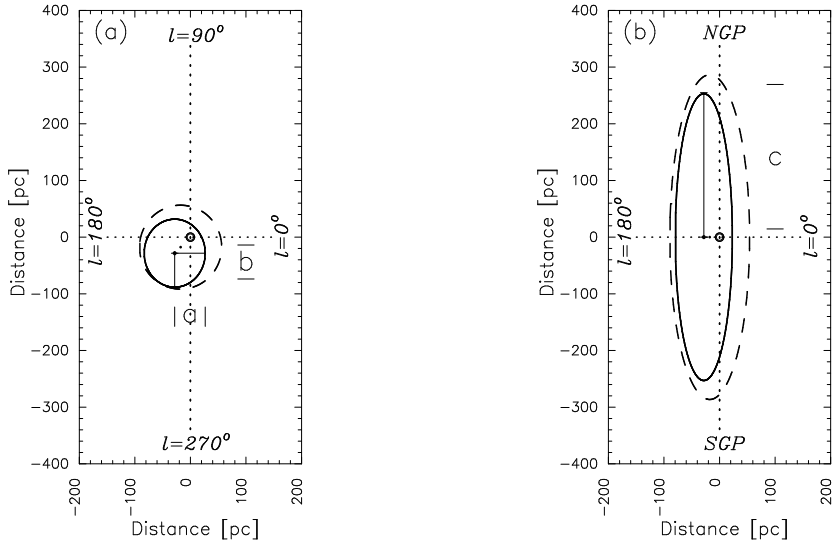,rheight=8.50cm,height=10.5cm,angle=270}}
}
\caption[]{
The geometry of the best fit model (Model II(b)) for the local scattering structure.
(a) is the section in the galactic plane.
(b) is the section along a plane perpendicular to the galactic plane and passing
through the north and the south galactic poles.
The solid geometry is for $ r_c \sim $ 35 pc, $a \sim $ 60 pc and $b \sim $ 60 pc,
and the dashed one has $ r_c \sim $ 20 pc, a $ \sim $ 75 pc and  b $ \sim $ 75 pc.
}
\label{Fig11}
\end{figure}

A unique determination of all parameters has not been possible from the present
observations. 
However, it has enabled us to get some useful insights on a possible scattering structure and
derive reasonable constraints on its size, the location and the strength of scattering. 
Note that the 
constraints on the semi-major axis c, the direction ($ l_c , b_c $ ) and the integrated strength of 
scattering from the shell are rather tight as these parameter values are critical in determining the 
scattering anomalies. 
However, we have not been able to derive similar tight constraints on rest of the parameters. 
Our present sample does not consist of many pulsars whose scattering properties are predominantly determined
by either the cavity only or the ambient ISM only and therefore our constraints on 
their strengths of scattering 
($ \overline C ^2 _{n,in} $ and $ \overline C ^2 _{n,out} $) are not very tight. 
Also, the 
present measurements do not allow us to put unique constraints on the size in the galactic plane 
(semi-major axes a and b) and the offset to the centre ($ r_c $) from the Sun. Our results can also be 
explained by a structure that is relatively bigger ($ a ~ \approx ~ b ~ \sim $ 75 pc), if
its centre is located nearer to the Sun ($ r_c ~ \sim ~ $ 20 pc). 
This is shown as the dashed geometry in
Figs. 11.a and 11.b. It can be stated that intermediate structures within $ 35 > r_c > 20 $ pc, 
$ 60 < a < 75 $ pc and $ 60 < b < 75 $ pc are also equally possible, with an asymmetry that becomes less
pronounced with the increase of size. It is not possible to resolve this ambiguity using the present
measurements alone.

\section{Discussion}

\subsection{Successes and limitations of the model }

We have attempted to explain our observations by considering a inhomogeneous distribution of electron density
fluctuations in the LISM. 
We have confined ourselves to the simplest possible density structures required to reproduce the observed 
trends.
Simple models in which the solar neighborhood has an enhanced or a reduced scattering strength relative 
to the ambient medium, fail to explain our observations. 
We need a three-component medium consisting of a shell of enhanced scattering surrounding the 
solar neighborhood which has lower scattering strength compared to the normal ISM.
The salient features of our model, shown schematically in Fig. 11, are:

(i) The scattering structure has an ellipsoidal morphology, with a size in the plane perpendicular to the
${\rm  b = 0^o }$ plane being about 5 times larger than in the galactic plane.

(ii) The Sun is located away from the centre of the structure, but must be well inside of it. 

(iii) The density fluctuations in the shell are much larger than those in the interior and in
the normal ISM. 
For nearby pulsars, the shell contributes substantially to the scattering as compared to the 
cavity and the normal ISM. 

(iv) There is a suggestion that the strength of scattering of the shell decreases 
with the height above the galactic plane.

(v) The strength of scattering in the inner cavity is about 5 times lower than that of the normal ISM.

This three-component model is able to reproduce a number of the observed trends in the data.
The model accounts for the observed scattering anomalies with good success (Figs. 9.k to 9.n).
It also accounts for the enhanced scattering of nearby pulsars and reproduces the observed trend 
in the variation of the decorrelation bandwidths with distance (Figs. 10.e and 10.f).
From our model, we are able to obtain constraints on the geometrical parameters and 
the scattering properties of the structure (Table 4).

We have investigated only the simplest models and cannot rule out the possibility of better fits 
with more complex models.
Our modelling is constrained by the limited number of pulsars used for probing the LISM.
The model has 9 free parameters and the present data are not good enough to constrain all of them
equally well.
We are able to put tight constraints on the size along the plane perpendicular to the galactic 
plane, the direction of the centre and the strength of scattering of the shell (Table 4). 
However, the strengths of scattering of the cavity and the ambient medium are poorly constrained.
Also, we are unable to distinguish between the small highly asymmetric structure and
a less asymmetric, but bigger structure (Fig. 11). 
Intermediate structures within $ 35 > r_c > 20 $ pc, $ 60 < a < 75 $ pc and $ 60 < b < 75 $ pc 
are also consistent. 
However, it is clear that the smaller structures need to be more asymmetric.
We cannot constrain the thickness of the shell since there seem to be no pulsars situated in the shell,
but can only estimate the integrated strength of scattering. 

\subsection{Implications for the electron density distribution in the LISM}
The electron density distribution in the LISM can be studied using the dispersion properties of nearby 
pulsars.
A detailed study is, however, not feasible at present because independent distance estimates are available 
only for two of our pulsars (Table 1).
The pulsar PSR B0950+08 that has a parallax distance $ \approx $ 130 pc (Gwinn et al. 1986) is located, 
in our model, within the cavity.
Its DM implies $ n_e ~ \approx ~ 0.02 ~ {\rm cm ^{-3}} $ which is about 4 times larger than its value 
based on the model given by Snowden et al. (1990) for the distribution of the soft X-ray background.

Scintillation measurements provide us information on the distribution of electron density fluctuations.
However, the apparent lack of understanding about the exact relations between the scattering strengths and 
other properties of the medium prevents us from deducing the $ n_e $ distribution directly from 
that of density fluctuations. 
The presence of an inhomogeneous medium is likely to be more prominent in scintillation owing to the 
nonlinear relation of scintillation properties to the electron density.
Therefore, even the regions of density enhancements that do not contribute significantly to the dispersion, 
are likely to be detectable through their scintillation effects. 
The Vela pulsar (PSR B0833-45) is a good example where a density enhancement of a factor of 5 
and a $ C_n^2 $ enhancement of a factor of $ \ga $ 100 are known to exist (Rickett 1977).
Also, unlike dispersion, scintillation effects are sensitive to the locations of the scattering regions, 
and therefore are useful to infer about the locations of regions of enhanced scattering.
Thus the information provided by the dispersion and the scintillation measurements appear to be complementary
and can enable us to derive much better insights on the $ n_e $ distribution in the LISM.

It is generally believed that the regions of enhanced electron densities are also the sites of enhanced 
electron density fluctuations ($ \Delta n_e ~ \propto ~ n_e $) (Rickett 1970). 
Therefore, an ellipsoidal shell structure for the density fluctuations would imply the existence 
of a dense electron density shell surrounding a low electron density cavity.
Our simplified model constrains only $ \int C_n^2 ~ dl $ and not the thickness of the shell, 
which is essential for determining the electron density in the shell. 
If we assume that $ C_n^2 ~ \propto ~ n_e^2 $, our observations imply a density contrast 
$ \sim $ 10$-$20 times between the shell and the ambient ISM in the case of a thin shell (d $ \sim $ 1 pc)
and $ \sim $ 5$-$8 times for $ \sim $ 10 pc thick shell.
The pulsar PSR B0823+26 lies well outside the bubble in our model and has a parallax distance 
of $ \approx $ 380 pc (Gwinn et al. 1986).
The shell is located at $ \sim $ 100$-$110 pc in the direction of this pulsar.
If we assume $ n_e ~ \approx ~ 0.02 ~ {\rm cm ^ {-3} } $ for the bubble interior and a canonical 
$ n_e $ of $ \approx $ 0.025 $ {\rm cm ^ {-3} } $ for the outer medium, this would mean 
a density contrast of $ \sim $ 30 for $ \sim $ 10 pc thick shell.
This is however $ \sim $ 3$-$6 times larger than that is estimated from our model.
Nevertheless, a possible density enhancement is indicated.

\begin{figure}[thbp]
\centerline{ \hskip 0 cm
{\psfig{file=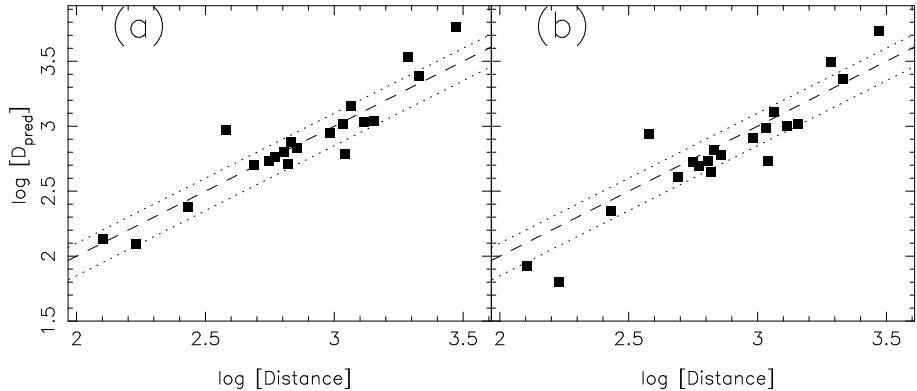,rheight=6.5cm,height=10.5cm,angle=270}}
}
\caption[]{
The predicted distance estimates ($ {\rm D _{pred} } $) of pulsars for a shell
density structure are plotted against their presently available distance
estimates. (a) is for the thin (thickness $ \sim $ 1 pc) shell of density
contrast of 10 and (b) is for the thick ($ \sim $ 10 pc) shell with a density
contrast of 8.The dashed lines correspond to a discrepancy of 30\%.
}
\label{Fig12}
\end{figure}

The distance estimates used in our analysis are based on the model (Taylor \& Cordes 1993) which does 
not take into account a possible inhomogeneous distribution of $ n_e $ in the LISM.
In order to examine the role of the LISM in determining the pulsar distances, we re-estimated them by 
incorporating a shell density structure for the LISM. 
We studied the cases of both the thin and the thick shells with the expected density contrasts estimated from 
the parameters of our scattering structure.
For these calculations, we have assumed $ n_e ~ \approx ~ 0.02 ~ {\rm cm ^ {-3} } $ for the inner cavity and 
the canonical $ n_e $ of $ \sim $ 0.025 $ {\rm cm ^ {-3} } $ for the ambient ISM.
We have also taken into consideration smooth variations of the densities in the shell and in 
the ambient medium with z-height.
Here we assumed an exponential decrease with a scale height $ \sim $ 500 pc for $ n_e $ in the ambient 
medium and a gaussian with a scale height $ \sim $ 125 pc for $ n_e $ in the shell. 

In the case of the thin shell with a density contrast of 10, we find the new distances to be similar to 
their presently available estimates for most pulsars (Fig. 12.a).
Barring a few anomalous ones (PSR B0823+26, PSR B2016+28, PSR B1508+55 and PSR B0919+06), the new distances
differ typically by 10\%.
For a higher density contrast of 20, the difference is only marginal.
In the case of the thick shell, dispersion due to the shell becomes more significant, 
but is still not enough to produce considerable changes in the distance estimates.
Discrepancies between the two distances are larger for a thick shell and they differ typically by 25\% for 
a 10 pc thick shell with a density contrast of 8 (Fig. 12.b). 
Therefore the density enhancements implied by our model 
do not alter the distance estimates to our pulsars 
significantly.
This gives a post-facto justification of our analysis procedure where we have used the available 
distance estimates and also suggests that a linear relation between the 
DMs and the distances is a reasonable approximation in the LISM despite its seemingly inhomogeneous 
nature of $ n_e $ distribution.
However, to establish this firmly, 
we need more accurate distance estimates for local pulsars than are available from 
the model of Taylor \& Cordes (1993). 

\subsection {Possible tests of the present model}

Observations of more pulsars will be useful for further constraining the parameters 
of our scattering model. 
The number of local pulsars (within a distance $ \la $ 1 kpc) has nearly doubled since we started our 
observations at Ooty as a result of recent pulsar surveys and forms a large enough sample to study the
$ n_e $ distribution in the LISM. 
Scintillation measurements of pulsars, however, require long-term observations to get reliable estimates 
of their strengths of scattering.
But there are simpler tests to verify the expectations based on our model and we discuss a few of them
here.
In particular, it is possible to use some of the recently discovered local pulsars to verify the 
predictions on their scattering properties. 
One example in this regard is PSR J1730$-$2304 ($ l ~ \approx ~ 3.1^o ~, ~ b ~ \approx ~ 6^o $), 
a pulsar located at closeby distance ($ \sim $ 510 pc), for which preliminary 
observations from the ORT shows that its enhanced scattering is consistent 
with model predictions.

In the specific cases of pulsars for which the scattering is due predominantly to the shell, 
one can treat the scattering, to a first order, as due to an equivalent thin screen located at the 
shell. 
In such cases, it is possible to use some distance-dependent scattering properties to check
the predicted locations of the shell boundaries in those directions. 
A straightforward test is the comparison between the proper motion and the scintillation speeds. 
The former ones are determined by pulsar distances whereas the latter ones also depend on the 
relative location of the screen (Gupta et al. 1994). 
The comparison, however, requires a careful analysis 
since the scattering geometries of our pulsars are represented by 
multi-component, inhomogeneous media.

Pulse broadening resulting from scattering is another property that is sensitive to the nature of 
distribution of the scattering material along the line-of-sight and hence can be used to distinguish 
between the thin screen and the extended screen geometries of scattering (Lyne \& Smith 1990;
Williamson 1974). 
The broadening is, however, not significant at 327 MHz for low DM pulsars and cannot be studied 
using our data. 
However, observations at lower frequencies can provide useful information.
Some promising pulsars in this aspect are PSR B1929+10, PSR B1133+16, PSR B0823+26 and PSR B2327$-$20 
where a broadening typical of a thin screen can be expected as per our model. 
In the case of pulsars PSR B1237+25 and PSR B0950+08, broadening features typical of those due to extended 
media are expected.
The observations, however, require high time resolution ($ \sim $ 0.1 msecs) and need to be carried 
out at observing frequencies $ \la $ 100 MHz for meaningful results. 

The data from observations at 25 MHz (Phillips \& Wolszczan 1990) reveal that 3 of our pulsars 
$-$ PSR B0823+26, PSR B0834+06 and PSR B0919+06 $-$ display pulse shapes with sharp rising edges
(Fig. 2 in Phillips \& Wolszczan (1990)).
For these pulsars, according to our model, scattering is predominantly ($ \ga $ 75\%) due to the 
shell material and hence the scattering geometry can be considered to be closer to that of a thin screen,
in which case broadening features characterized by a sharp rising edge followed by an exponential tail 
can be expected.
Given the signal-to-noise ratio of the profile of PSR B1604$-$00, such a feature is not apparent and
the pulse shape seems to imply a diffused scattering medium in its line-of-sight.
For this pulsar, our model predicts substantial amounts of scattering due to the shell material as
well as the outer ISM and, therefore, the scattering geometry can be treated more like that of 
a `continuous medium', which is in accord with the observed pulse shape.
Pulse shapes of PSR B0950+08 and PSR B1133+16 are hard to interpret since their profiles seem to
be changing significantly with the observing frequency and the scattering is low.
Leaving out these two pulsars, pulse shapes of the remaining 4 are
consistent with expectations of our scattering model.
Further low frequency observations of nearby pulsars will be useful.

Decorrelation bandwidth is another observable that depends on the location of the scattering screen. 
For a given location of the screen, it is essentially determined by the angular broadening of the scattered 
rays.
Therefore, simultaneous measurements of these two quantities will be useful to derive the effective location 
of the scattering screen and for pulsars with predominant scattering from the shell, it essentially means
the location of the shell. 
Angular broadening measurements, therefore, also form suitable techniques to test our scattering model. 
Although such observations have been reported recently (Gwinn, Bartel \& Cordes 1993) for 10 pulsars, 
the scattering disks of 4 common pulsars $-$ PSR B1919+21, PSR B1929+10, PSR B2016+28 and PSR B2020+28,
were unresolved and, therefore, interpretation of their angular broadening measurements is uncertain.
Pulsars which show relatively much higher levels of scattering, 
such as PSR B0823+26, PSR B1133+16 and PSR B$2327-20$, 
form some of the suitable candidates for further observations.

\subsection{Comparison with other studies of the LISM }

Snowden et al. (1990) have derived the three-dimensional geometry for the X-ray emitting cavity using their
`displacement model' technique. 
They have considered two specific models, which are the cases of a cavity formed (i) by simply removing 
the material and (ii) by sweeping out the material to form a shell at the edge of the emission region. 
The local scattering structure deduced by us has a shell morphology and favors this second model, if we 
assume $ \Delta n_e ~ \propto ~ n_e $. 
The size of our ellipsoidal shell is quite similar to their larger cavity, 
(Figs. 6 and 7 in Snowden et al. (1990)) suggesting a connection between the scattering structure and the 
X-ray cavity. 
The inferred cavity is much more extended away from the galactic plane with a striking asymmetry between
the northern and the southern galactic hemispheres (Cox \& Snowden 1986; Snowden et al. 1990).
The scattering structure derived by us also has a morphology that is much more extended away from the 
galactic plane and therefore resembles to that of the X-ray cavity.
However, an asymmetry between the two hemispheres has not been clearly established for our scattering
structure (see the constraints on location of the centre given in Table 4). 
The present data are not good enough to draw a firm conclusion on this.
Also, we find that the Sun needs to be significantly off-centred within the structure, a property which 
was not clearly established in the X-ray studies owing to regions of data with non-X-ray contamination 
and of poor spatial coverage. 

Warwick et al. (1993) studied the properties of EUV source population using the data from ROSAT. 
Their simple model of a spherical bubble of average radius $ \sim ~ 100 \pm 25 $ pc centred at the 
Sun with an interior gas density $ \sim $ 0.05 ${\rm  cm ^ {-3} } $ can account for the observed 
characteristics of the global source counts.
The bubble radius derived from their analysis and the radius of an equivalent sphere for our 
scattering structure are comparable suggesting a zeroth order consistency between the two. 
They also studied the spatial distribution of the sources and identified 
anomalous regions with a deficit or an excess compared to the global average, which they interpreted 
in terms of deviations from their `fiducial' bubble model.
In particular, they estimated the distances to the bubble boundaries towards some specific regions 
where prominent features were seen.
They suggested the existence of a relatively nearby ($ \sim $ 10 pc) absorbing wall towards the general 
direction of the Galactic Centre to explain an observed deficiency of a factor of 5 in the number of 
sources.
A closeby location ($ \sim $ 20 pc) of the shell in this direction is expected from our scattering 
structure too.
An excess of a factor of 2 was seen in the southern part of the Galactic quadrant III 
($ l ~ \approx ~ 200 ^o ~ , ~ b ~ \approx ~ - 30 ^o $), from which they inferred a bubble extent of 
$ \sim $ 100 pc, which is comparable to the distance to the shell boundary estimated from our model
($ \sim $ 100$-$110 pc) in this direction.
Also, from an excess in the northern sky ($ l ~ \approx ~ 120 ^o ~ , ~ b ~ \approx ~ 45 ^o $),
it was suggested that the bubble extends well beyond $ \sim $ 120 pc towards this region.
Given the uncertainties of our model parameters, a bubble extent as large as $ \sim $ 80 pc can be 
expected in this direction.
Therefore, it appears that the morphology and the size of our scattering structure are in broad 
agreement with those of the UV bubble suggesting a possible connection between the distributions of 
electron density fluctuations and that of absorbing gas in the LISM. 

Hajivassiliou (1992) has proposed an ellipsoidal envelope of high plasma turbulence around the 
Sun to explain the large scale features seen in the turbulence map derived using angular source 
size measurements from an interplanetary scintillation (IPS) survey made at 81.5 MHz 
(Fig. 1 in Hajivassiliou (1992)). 
This interpretation requires the Sun to be lying at the edge of the envelope in the direction
$ l ~ \approx ~ 30^o $ to explain the observed directional anisotropy of the turbulence parameter. 
Although, the scattering structure deduced by us has a similar morphology, there is little agreement 
on other properties such as the size, the location and the strength of scattering.
The estimates on the size of our scattering structure are much larger than that of the ellipsoidal
envelope suggested from the IPS data.
The morphology and the size of the envelope were based on the X-ray data (Cox \& Snowden 1986), where 
radii of $ \sim $ 150 pc and $ \sim $ 45 pc were typical for the ellipsoidal sections perpendicular to the
galactic plane and in the plane respectively which are much smaller than those inferred from our data 
(radii of $ \sim $ 250 pc and $ \sim $ 60 pc on these planes) (Fig. 11).
Also, we find in our model the Sun needs to be located well inside the shell structure (offset $ \sim $ 
20$-$35 pc) to explain the observed scattering anomalies of pulsars.
Considerable discrepancy exists in the estimates of strengths of scattering of the shell, where the 
integrated strength of scattering constrained from our observations (Table 4) is $ \sim $ 2$-$5 times
smaller than that was required to explain the IPS data.
Furthermore, in our model, scattering due to the shell decreases with the height above the galactic plane
and becomes an order of magnitude lower at $ | b | ~ \approx ~ 90 ^o $,
something that has not been considered by Hajivassiliou (1992).

Studies of the distribution of the neutral hydrogen in the LISM (Frisch \& York 1983; Paresce 1984)
reveal the existence of a large region, of $ \sim $ 100 pc surrounding the Sun, that is virtually devoid of
neutral hydrogen (density $ {\rm n_H } ~ \la ~ 0.1 ~ {\rm cm ^ {-3} } $).
This `void' has an opening in the Galactic quadrant III (Fig. 4 in Paresce (1984)).
The Sun seems to be located away from the centre of this void and a boundary, defined as the distance 
where the column density of neutral hydrogen becomes $ \sim ~ 10 ^{19} ~ {\rm cm ^{-2} } $ 
(Cox \& Reynolds 1987), is fairly closeby in the Galactic quadrant I (towards $ l ~ \approx ~ 45 ^o $)
whereas it is at about 50 pc away from the Sun in the quadrants II and IV. 
It is interesting to note that in the quadrants I, II and IV, the morphology of our scattering structure
has a broad resemblance to that of this HI void.
Also our estimates to the bubble boundaries in the quadrants I, II and IV are comparable to those of the
HI void.
In our model, the bubble region extends to much larger distances in the quadrant III. 
A similar property can be seen in the HI void, suggesting a broad consistency between the two.
On the basis of the morphological similarity between the two structures, there appears to exist a 
probable connection between the interior of our shell structure and the HI void.

In view of the morphological agreement between the X-ray emitting cavity, the UV bubble, the neutral 
hydrogen void and the local density structure from our observations, properties of the large-scale 
distributions of the neutral and the ionized materials in the Local Bubble can be summarized as follows.
The hot gas giving rise to the soft X-ray emission seem to fill the local void of neutral hydrogen.
The deficiency of neutral hydrogen enhances the chances of detections of the EUV sources, which is
confirmed from the recent observations. 
Also, the observed properties of EUV sources and their spatial distribution are broadly consistent with 
the morphology of the void.
Lower magnitudes of electron density fluctuations seem to prevail in this region, which is broadly 
consistent with lower electron densities expected from the displacement model.
Our observations also suggest that the bubble region is surrounded by a shell of much higher density
fluctuations.
This implies a similar structure for the distribution of $ n_e $, which needs to be confirmed.

\section{Conclusion}

For the first time, the structure of the LISM has been modelled using the results from a 
systematic, long-term pulsar scintillation study. 
Our analysis based on the scintillation properties of twenty nearby pulsars suggests that the large-scale
distribution of the ionized material in the solar neighborhood is not uniform.
Systematic trends have been seen in the scattering properties of pulsars, which imply a coherent 
structure of electron density fluctuations in the LISM.
The detailed analysis of the observed anomalous scattering effects shows that such a structure is highly 
asymmetric relative to the location of the Sun.
Simple models in which the solar neighborhood has an enhanced or a reduced scattering strength relative to 
the ambient medium fail to reproduce the scattering anomalies.
To explain our observations, we need a three-component scattering medium in which the solar neighborhood 
is surrounded by a shell of much higher density fluctuations embedded in the normal, large scale ISM.
We are also able to put reasonable constraints on the geometrical and the scattering properties
characterizing the size, location and density fluctuations of such a structure.
The shell has an ellipsoidal morphology and is much more extended away from the galactic plane
than in the plane, with radii of $ \sim $ 270$-$330 pc and $ \sim $ 60$-$75 for the sections along the
galactic poles and through the galactic plane respectively.
The Sun is located away from the centre by about $ \sim $ 20$-$35 pc.
The density fluctuations in the shell are much larger than those in the interior and in the outer region
and there is a suggestion that it decreases with the height above the galactic plane.
We also find that the morphology of our scattering structure is similar to that of the Local Bubble 
known from various earlier studies based on HI, X-ray and UV data.
The LISM and the distribution of electron density fluctuations in it are likely to be more complex than
that is suggested by our simplified model, but we hope the present work will serve as a useful framework 
within which more detailed questions can be addressed.

{
{\it Acknowledgments:}
The authors would like to thank J. Chengalur and M. Vivekanand for reading the manuscript 
and giving useful comments.
We also thank M. Vivekanand for providing the software for pulsar data acquisition, 
and V. Balasubramanian for the telescope time and technical help with the observations.
We thank the referee, S. R. Spangler, for several fruitful comments and suggestions, which 
improved the contents of the paper.
}


\newpage

\addtolength{\oddsidemargin}{-1cm}
\addtolength{\evensidemargin}{-1cm}

\begin{table}
\caption{\sc The pulsar sample and the observing parameters}
\vspace{0.5cm}
\begin{tabular}{rcrrrrlrll}
\hline
\hline
   &                  &      &        &     &     &                     &     &         &       \\
No. & Pulsar & DM & D & $ l $ & $ b $ &  Period of &  $ {\rm N_{ep} } $ & 
$ {\rm \Delta f _{ch} } $ & $ {\rm \Delta t } $ \\
& & (${\rm{pc~cm^{-3}}}$) & (pc)& (deg) & (deg) & observation & & (kHz) & (secs) \\
   &                  &      &        &     &     &                     &     &         &       \\
\hline
\hline
   &                  &      &        &     &     &                     &     &         &       \\
 1 & PSR \ B$0031-07$ & 10.9 &  680   & 110 &$-70$&  April 95$-$July 95 &   2 &$140.60$	&$9.43$	\\
 2 & PSR \ B$0329+54$ & 26.8 & 1430   & 145 &$- 1$&  April 95$-$July 95 &  14 &$140.60$ 	&$7.15$	\\
 3 & PSR \ B$0628-28$ & 34.4 & 2140   & 237 &$-17$&  October 93$-$January 94 &  17 &$140.60$	&$12.44$\\
 4 & PSR \ B$0823+26$ & 19.5 &$380^@$ & 197 &$ 32$&  March 93$-$January 94 &  31 &$140.60^\dagger$&$13.27$\\
 5 & PSR \ B$0834+06$ & 12.9 &  720   & 220 &$ 26$&  January 93$-$July 95 &  93 &$140.60^\dagger$	&$12.74$\\
 6 & PSR \ B$0919+06$ & 27.2 &$>$2970 & 225 &$ 36$&  March 94$-$June 94 &  19 &$140.60$	&$21.53$\\
 7 & PSR \ B$0950+08$ &  3.0 &$130^@$ & 229 &$ 44$&  October 93$-$July 95 &   3 &$281.20$	&$12.65$\\
 8 & PSR \ B$1133+16$ &  4.8 &  270   & 242 &$ 69$&  February 93$-$July 95 &  59 &$140.60^\dagger$	&$11.88$\\
 9 & PSR \ B$1237+25$ &  9.3 &  560   & 253 &$ 87$&  October 93$-$January 94 &   9 &$140.60$	&$13.82$\\
10 & PSR \ B$1508+55$ & 19.6 & 1930   &  91 &$ 52$&  April 95$-$July 95 &   9 &$140.60$	&$14.79^\ddagger$\\
11 & PSR \ B$1540-06$ & 18.5 & 1160   &   1 &$ 37$&  April 95$-$July 95 &  12 &$140.60$	&$14.18$\\
12 & PSR \ B$1604-00$ & 10.7 &  590   &  11 &$ 36$&  April 95$-$July 95 &  10 &$140.60$	&$21.09$\\
13 & PSR \ B$1747-46$ & 21.7 & 1080   & 345 &$-10$&  April 95$-$July 95 &  12 &$140.60$	&$14.85^\ddagger$\\
14 & PSR \ B$1919+21$ & 12.4 &  660   &  56 &$  4$&  March 93$-$January 94 &  63 &$140.60^\dagger$	&$13.37	$\\
15 & PSR \ B$1929+10$ &  3.2 &  170   &  47 &$- 4$&  March 94$-$June 94 &   9 &$140.60$	&$11.33$\\
16 & PSR \ B$2016+28$ & 14.2 & 1100   &  68 &$- 4$&  October 93$-$January 94 &  20 &$140.60$	&$13.95$\\
17 & PSR \ B$2020+28$ & 24.6 & 1300   &  69 &$- 5$&  March 94$-$June 94 &  15 &$140.60$	&$17.20^\ddagger$\\
18 & PSR \ B$2045-16$ & 11.5 &  640   &  31 &$-33$&  October 93$-$January 94 &  35 &$140.60$	&$19.62$\\
19 & PSR \ B$2310+42$ & 17.3 &  960   & 104 &$-16$&  April 95$-$July 95 &  10 &$140.60$	&$17.47$\\
20 & PSR \ B$2327-20$ &  8.4 &  490   &  49 &$-70$&  March 94$-$June 94 &  18 &$140.60$	&$32.87$\\
   &                  &      &        &     &     &                     &     &         &       \\
\hline
\hline
   &                  &      &        &     &     &                     &     &         &       \\
\end{tabular}

{$^@$} {Distance estimates from parallax method (Gwinn et al. 1986).}

{$^{\dagger}$} {Part of the data were taken with $ \Delta f _{ch} $ = 281.20 kHz.}

{$^{\ddagger}$} {Part of the data were taken with larger $ \Delta t $ (twice the value given here).}

\end{table}

\clearpage

\begin{table}
\caption{\sc The measured scintillation parameters and the \cn estimates}
\vspace{0.5cm}
\begin{tabular}{crccrcrcrcrc}
\hline
\hline
  &   &               &  &               &  &               & &                    & &          &  \\
 & No. & Pulsar        & & $ \nu _{d,g} $ & & $ \tau _{d,g} $ & & log  \cn & & log  \cn & \\
   &   &          & & (kHz) & & (secs) & &   & & (CWB)$^\dagger $ & \\
  &   &               &  &               &  &               & &                    & &          &  \\
\hline
\hline
  &   &               &  &               &  &               & &                    & &          &  \\
  &  1& PSR B$0031-07$&  & 1039$\pm $208 &  &  2961$\pm $592& &  $-4.22 \pm 0.058$ & &          &  \\
  &  2& PSR B$0329+54$&  &    165$\pm $13&  &  307$\pm $25  & &  $-4.03 \pm 0.024$ & &  $-3.46$ &  \\
  &  3& PSR B$0628-28$&  &    203$\pm $18&  &  455$\pm $41  & &  $-4.54 \pm 0.054$ & &  $-4.48$ &  \\
  &  4& PSR B$0823+26$&  &    293$\pm $41&  &  126$\pm $19  & &  $-3.24 \pm 0.025$ & &  $-3.01$ &  \\
  &  5& PSR B$0834+06$&  &    454$\pm $27&  &  390$\pm $23  & &  $-3.91 \pm 0.030$ & &  $-3.33$ &  \\
  &  6& PSR B$0919+06$&  &    256$\pm $41&  &  160$\pm $27  & &  $-4.82 \pm 0.023$ & &  $-3.80$ &  \\
  &  7& PSR B$0950+08$&  &   $\gg $9000  &  & $>$7500       & &  $ \ll -3.70     $ & &  $-2.91$ &  \\
  &  8& PSR B$1133+16$&  &    816$\pm $57&  &  165$\pm $12  & &  $-3.32 \pm 0.052$ & &  $-3.31$ &  \\
  &  9& PSR B$1237+25$&  &   1828$\pm$128&  &  439$\pm $26  & &  $-4.07 \pm 0.085$ & &  $-3.97$ &  \\
  & 10& PSR B$1508+55$&  &    197$\pm $33&  &  158$\pm $30  & &  $-4.40 \pm 0.017$ & &  $-4.13$ &  \\
  & 11& PSR B$1540-06$&  &    111$\pm $12&  &  526$\pm $58  & &  $-3.73 \pm 0.026$ & &          &  \\
  & 12& PSR B$1604-00$&  &    378$\pm $19&  &  933$\pm $56  & &  $-3.71 \pm 0.057$ & &  $-3.66$ &  \\
  & 13& PSR B$1747-46$&  &    165$\pm $21&  &  215$\pm $30  & &  $-3.85 \pm 0.028$ & &          &  \\
  & 14& PSR B$1919+21$&  &    285$\pm $14&  &  374$\pm $22  & &  $-3.95 \pm 0.060$ & &  $-2.84$ &  \\
  & 15& PSR B$1929+10$&  &   1293$\pm $78&  &  348$\pm $24  & &  $-3.13 \pm 0.082$ & &  $-2.94$ &  \\
  & 16& PSR B$2016+28$&  &    206$\pm $12&  &  995$\pm $60  & &  $-3.97 \pm 0.053$ & &  $-3.40$ &  \\
  & 17& PSR B$2020+28$&  &    270$\pm $22&  &  279$\pm $22  & &  $-4.10 \pm 0.020$ & &  $-3.74$ &  \\
  & 18& PSR B$2045-16$&  &    539$\pm $65&  &  138$\pm $17  & &  $-3.92 \pm 0.018$ & &  $-3.81$ &  \\
  & 19& PSR B$2310+42$&  &    114$\pm $15&  &  309$\pm $40  & &  $-3.46 \pm 0.022$ & &          &  \\
  & 20& PSR B$2327-20$&  &    268$\pm $11&  &  432$\pm $13  & &  $-3.41 \pm 0.039$ & &          &  \\
  &   &               &  &               &  &               & &                    & &          &  \\
\hline
\hline
  &   &               &  &               &  &               & &                    & &          &  \\
\end{tabular}

{$^{\dagger}$}{Measurements of \cn from Cordes et al. (1985) corrected for the new pulsar distance estimates.}
\end{table}

\clearpage

\begin{table}
\caption{\sc Models for the local scattering structure}
\vspace{0.5cm}
\begin{tabular}{cccllll}
\hline
\hline
	    &             &                                                             & & & &\\
 Class & Model & Description & $ \epsilon _A $ &  $ \epsilon _A $ & $ \epsilon _B $ & Remarks \\ 
       &       &             & (DM)            &  (D)             &                 &         \\
	    &             &                                                             & & & &\\
\hline
\hline
	    &             &                                                             & & & &\\
2-component & Model I(a) & cavity of lower density fluctuations & 0.0937 & 0.0293 & 0.3543 & fails to explain \\
            &             & $ C ^2 _{n,in} < C ^2 _{n,out} $       & & &                & the observations \\
            &             &                                          & & &              &  \\
            & Model I(b) & cavity of higher density fluctuations & 0.0485 & 0.0442 & 0.2146 & fails to explain \\
            &             & $ C ^2 _{n,in} > C ^2 _{n,out} $       & & &                & the observations \\
            &             &                                          & & &                &\\
3-component&Model II(a)&shell of enhanced density fluctuations & 0.0604$^\dagger $ & 0.0158$^\dagger $ & 
0.1429$^\dagger $&reasonable \\
            &  & $ C ^2 _{n,sh} \gg C ^2 _{n,in} $    & 0.1257$^\ddagger $ & 0.0172$^\ddagger $ & 
0.1092$^\ddagger $ &agreement\\
            &  & $ C ^2 _{n,sh} \gg C ^2 _{n,out} $       & & &                &\\
            &             &                                          & & &                &\\
            & Model II(b) & similar to Model II(a)    & 0.0099$^\dagger $ & 0.0073$^\dagger $ & 0.1189$^\dagger $ & best \\
            &    & with an additional feature of & 0.0191$^\ddagger $ & 0.0109$^\ddagger $ & 0.0779$^\ddagger $ & agreement \\
            &             & $|z|$-dependent scattering for the shell & & &           &     \\
            &             & $ C ^2 _{n,sh} (z) $ has a scale height $ z_d \sim $ 135 pc & & & &\\
	    &             &                                                             & & & &\\
\hline
\hline
	    &             &                                                             & & & &\\
\end{tabular}

{$^{\dagger}$ } {For the solid geometry shown in Fig. 11}

{$^{\ddagger}$ } {For the dashed geometry shown in Fig. 11}

\end{table}

\clearpage

\begin{table}
\caption{\sc Geometrical parameters and scattering properties of the best fit model}
\vspace{0.5cm}
\begin{tabular}{lc}
\hline
\hline
                        &       \\
Model parameter		&	\\
                        &       \\
\hline
\hline
                        &       \\
Physical dimensions of the ellipsoid (semi-major axes {\it a}, {\it b} and {\it c}) :			\\
                                                                &                        \\
Perpendicular to the galactic plane	(NGP-SGP cut)	 	& 	$ 270 ~ < ~ c ~ < ~ 330 $ pc \\
In the galactic plane 			($ 0^o - 180^o $)$^{\dagger} $	& 	$  60 ~ < ~ a ~ < ~  75 $ pc \\
In the galactic plane			($ 90^o - 270^o $)$^{\dagger}$	& 	$  60 ~ < ~ b ~ < ~  75 $ pc \\
                                                                &                        \\
Location of the centre of the ellipsoidal shell : 					 \\
                                                                &                        \\
Galactic longitude 			&	$ 215^o ~ < ~ l_c ~ < ~ 240^o $		 \\
Galactic latitude			&	$ -20^0 ~ < ~ b_c ~ < ~ 20^o $		 \\
Offset from the Sun$^{\dagger} $		&	$  35   ~ > ~ r_c ~ > ~ 20 $ pc          \\
                                                                &                        \\
Strengths of scattering ($C_n^2$) of different components :			         \\
                                                                &                        \\
Inner cavity	&	$ 10 ^ {-4.70} ~ < ~ \overline {C_n^2} ~ < 10 ^ { - 4.22 } 
~  {\rm m ^ { - 20 / 3 }} $       \\
Shell material (with thickness {\it d})	& $ 10 ^ { -0.96 } ~ < ~ \int _0 ^d  ~ C_n^2 (l) ~ dl ~ < 
~ 10 ^ { - 0.55 } ~ {\rm pc ~ m ^ { - 20 / 3 }} $  \\
Outer ISM			&	\hspace{0.5cm} $ \overline {C_n^2} ~ < ~ 10 ^ { - 3.30 } 
~ {\rm m ^ { - 20 / 3 }} $	 \\	
                        &       \\
\hline
\hline
                        &       \\
\end{tabular}

{$^{\dagger}$} {Smaller values of {\it a} and {\it b} require a larger value 
of $ r_c $ to reproduce the results.}

\end{table}

\end{document}